\documentclass[letterpaper,12pt,linesnumbered,ruled,vlined]{article}
\usepackage{graphicx}
\usepackage{epstopdf}
\usepackage{amsmath}
\usepackage{epsfig}
\usepackage{float}
\usepackage[caption = false]{subfig}
\usepackage{amsfonts}
\usepackage{rotating}
\usepackage{fullpage}
\usepackage{setspace}
\usepackage[longnamesfirst, authoryear]{natbib}
\usepackage{hyperref}
\usepackage{enumerate}
\usepackage{booktabs,threeparttable}
\usepackage{authblk}
\usepackage{multirow}
\usepackage[english]{babel}
\usepackage[utf8]{inputenc}
\usepackage{tikz,,mathpazo}
\usepackage{setspace}

\usepackage{epigraph}

\usepackage{mathptmx}
\usepackage{color}
\usepackage{array}
\usepackage{float}
\usepackage{booktabs}
\usepackage{mathtools}
\usepackage{multirow}
\usepackage{amsmath}
\usepackage{amsthm}
\usepackage{amssymb}
\usepackage{graphicx}
\usepackage{rotfloat}

\makeatletter

\DeclareTextSymbolDefault{\textquotedbl}{T1}

\floatstyle{ruled}
\newfloat{algorithm}{tbp}{loa}
\providecommand{\algorithmname}{Algorithm}
\floatname{algorithm}{\protect\algorithmname}

\@ifundefined{date}{}{\date{}}

\date{\today}

\usepackage{amsfonts}
\usepackage{dsfont}

\allowdisplaybreaks[4]
\DeclareMathAlphabet{\mathcal}{OMS}{cmsy}{m}{n}

\usepackage{etoolbox}
\usepackage{authblk}

\makeatletter
\patchcmd{\epigraph}{\@epitext{#1}}{\itshape\@epitext{#1}}{}{}
\makeatother

\hypersetup{
	colorlinks,
	citecolor=blue,
	filecolor=black,
	linkcolor=blue,
	urlcolor=black
}

\newtheorem{definition}{Definition}

\mathchardef\mhyphen="2D

\allowdisplaybreaks

\begin{document}

	\begin{singlespace}
		\title{Optimal Trade and Industrial Policies in the Global Economy: A Deep Learning Framework\Large{\thanks{We are grateful to Yucheng Yang, Jeffrey Sun, Costas Arkolakis, Daniel Xu, Dan Lu, and numerous seminar participants for helpful suggestions and comments. The ideas expressed herein are solely our own and do not represent the views or opinions of any institution or organization.
		}}}
		\author[1]{Zi Wang\thanks{wangzi@hkbu.edu.hk}}
		\author[2]{Xingcheng Xu\thanks{xingcheng.xu18@gmail.com}}
		\author[3]{Yanqing Yang\thanks{yanqingyang@fudan.edu.cn}}
		\author[4]{Xiaodong Zhu\thanks{xdzhu@hku.hk}}
		\affil[1]{Hong Kong Baptist University}
		\affil[2]{Shanghai Artificial Intelligence Laboratory}
		\affil[3]{Fudan University}
		\affil[4]{The University of Hong Kong}
		
		\date{July 23, 2024}
		\maketitle
		\begin{abstract}
			\singlespace\noindent 
			\newline
			We propose a deep learning framework, \emph{DL-opt}, designed to efficiently solve for optimal policies in quantifiable general equilibrium trade models. \emph{DL-opt} integrates (i) a nested fixed point (NFXP) formulation of the optimization problem, (ii) automatic implicit differentiation to enhance gradient descent for solving unilateral optimal policies, and (iii) a best-response dynamics approach for finding Nash equilibria. Utilizing \emph{DL-opt}, we solve for non-cooperative tariffs and industrial subsidies across 7 economies and 44 sectors, incorporating sectoral external economies of scale. Our quantitative analysis reveals significant sectoral heterogeneity in Nash policies: Nash industrial subsidies increase with scale elasticities, whereas Nash tariffs decrease with trade elasticities. Moreover, we show that global dual competition, involving both tariffs and industrial subsidies, results in lower tariffs and higher welfare outcomes compared to a global tariff war. These findings highlight the importance of considering sectoral heterogeneity and policy combinations in understanding global economic competition.
						
			\vspace{0.2cm}						
		\end{abstract}
	\end{singlespace}
	\thispagestyle{empty}

	\section{Introduction\label{sec:Introduction}}

 In recent years, economic conflicts have arisen among major global economies, necessitating a comprehensive understanding of the incentives and consequences associated with these conflicts for both economic studies and policy-making. Within this context, three trends have become increasingly prominent. Firstly, countries are relying on a combination of trade and industrial policies to compete with one another. The most notable recent example is China’s export-promoting trade policies and Made-in-China 2025 initiatives that seek to use government subsidies to develop certain advanced technology sectors that are deemed essential to the future competitiveness of China’s manufacturing industry. By no means is China the only country that has utilized trade and industrial policies in recent years. Juhaz, Lane, and Rodrik (2023) have documented a steady rise of industrial policy interventions globally in the last decade and a half, and the pace of the increase rose significantly around 2018. 
 
Secondly, in many real-world scenarios, countries behave as if engaged in non-cooperative games, necessitating consideration of other countries’ potential retaliatory actions when formulating optimal policies. A prime example of this is the US government’s response to China’s trade and industrial policies, as documented by Ju, Ma, Wang, and Zhu (2023). To counter China’s Made-in-China 2025 initiatives, the Trump administration launched a tariff war against China in 2018. Later, the Biden administration further escalated the technology competition between the US and China by resorting to export control and its own industrial policies, as outlined in the CHIPS Act and Inflation Reduction Act.

Finally, countries have also realized that their economies are deeply interconnected through global value chains, and a complete decoupling would result in significant economic losses for all parties. In light of these considerations, both the US and EU have emphasized derisk rather than decoupling in managing their economic relationship with China.

Given these important trends, it is essential to evaluate the costs of economic conflicts and the benefits of potential economic cooperation by examining Nash equilibria in multi-country, multi-sector quantitative general equilibrium models that incorporate global input-output linkages and a combination of policies.

However, solving for high-dimensional continuous optimal policies and the corresponding Nash equilibria in quantitative general equilibrium trade models poses computational challenges. Three primary difficulties arise. First, conditional on a set of policies, equilibrium outcomes must be determined by solving a high-dimensional nonlinear system. Second, the policy space often includes numerous continuous policy tools. Third, finding Nash equilibria requires iteratively determining unilateral optimal policies for each country.
	
There are two main approaches in the literature to address the challenges above. First, \citet{Judd2012} develop the mathematical programming with equilibrium constraints (MPEC) framework. When combined with standard algorithms for constrained optimization, this approach offers a speed advantage because it avoids repeatedly solving the equilibrium system for each policy guess. However, this speed advantage diminishes when the dimension of equilibrium conditions is high due to the exploding dimensionality of the solution space.\footnote{For example, based on \citet{Judd2012}, \citet{Ossa2014} required four days to compute optimal tariffs (both unilaterally and mutually) in a general equilibrium model with 7 economies and 33 industries, but without input-output linkages, using a high-end desktop computer and standard MATLAB software. Incorporating more economies, industries, and policies, as well as input-output linkages, presents significant computational challenges.}
	
Second, \citet{Lashkaripour2019} and \citet{Bartelme2019} derive sufficient statistics of optimal policies without solving the high-dimensional optimization problem. This approach is appealing since it does not require computation power but can get optimal policies by calculating several simple and intuitive sufficient statistics. However, this approach has to make simplification assumptions to get the exact form of sufficient statistics.\footnote{For example, the analytical forms of unilaterally optimal policies in \citet{Lashkaripour2019} rely on the ``\emph{internal cooperation}" assumption: they assume that the relative wages in other economies remain unchanged under one economy's optimal policies.} As a result, their optimal policies are, in fact, constrained optimal policies under simplification assumptions. To obtain fully optimal policies without these assumptions, we still need to solve the high-dimensional nonlinear optimization problem.
	
	In this paper, we propose a deep learning framework, \emph{DL-opt}, for solving optimal (unilaterally and mutually) policies in quantifiable general equilibrium trade models. To avoid the exploding dimensionality of solution space in MPEC, we do not include equilibrium outcomes in the solution space and take equilibrium conditions as constraints. Instead, we employ the nested fixed point (NFXP) algorithm developed by \citet{Rust2000} that solves the equilibrium system by iteration at each guess of policies. Utilizing NFXP, our focus is on improving the efficiency of policy updating. To this end, we employ the widely-used deep learning algorithm that can efficiently update the policies based on the gradient of the objective function with respect to the parameters. The key to gradient descent is automatic implicit differentiation, which efficiently computes the gradient of equilibrium outcomes with respect to policies via the implicit function defined by the equilibrium system.
	
	Armed with our deep learning method to solve unilaterally optimal policies, we employ state-of-the-art best-response dynamics methods for solving Nash equilibria. In particular, we use a random-shuffle playing sequence to improve algorithm efficiency and avoid being trapped in local solutions. In sum, by combining the NFXP algorithm, automatic implicit differentiation, and best-response dynamics, our \emph{DL-opt} framework is able to solve for high-dimensional non-cooperative policies in general equilibrium trade models.
	
	We then apply our \emph{DL-opt} framework to solve optimal trade and industrial policies in a multi-country-multi-sector general equilibrium trade model with input-output linkages and sectoral scale economies, \`a la the models of \citet{Lashkaripour2019}, \citet{Bartelme2019}, and \citet{JMWZ2023}. We consider the world with 6 major economies plus the rest of the world (ROW) and 44 sectors (22 of them are tradable sectors). Using one CPU in a laptop, we compute Nash tariffs and industrial policies among 7 regions (6 major economies + ROW) in this model by about 5 hours. In comparable computational exercises, our method demonstrates an efficiency at least ten times greater than that of widely used nonlinear solvers, such as Knitro, particularly when analytical Jacobian and Hessian matrices are unavailable. Our solutions are shown to be precise, with a stable convergence process. Additionally, we demonstrate that parallelization can further enhance the efficiency of our method.
	
	Our quantitative explorations focus on \emph{global dual competition} in which each economy chooses import tariffs and industrial subsidies to maximize its own real income. To isolate the impacts of trade wars and understand the independence of trade and industrial policies, we also consider the \emph{global trade war} in which each economy chooses import tariffs to maximize its own real income. Nash policies computed using our \emph{DL-opt} framework shed light on global competition in the following aspects:
	
	First, Nash industrial subsidies increase with sectoral scale elasticities but are not correlated with trade elasticities. Consistent with previous studies such as \citet{Lashkaripour2019}, these results suggest that industrial subsidies can be used to address misallocation across sectors led by external economies of scale. The terms-of-trade effect, decreasing with trade elasticities, is not a primary concern in determining optimal industrial subsidies.
	
	Second, Nash tariffs under global dual competition increase with scale elasticities but decrease with trade elasticities. Moreover, we find that the positive correlation between Nash tariffs and scale elasticities is much stronger under trade wars than under dual competition. This result suggests that when industrial policies are not allowed, tariffs have to be used to address misallocation across sectors and thereby be prohibitively high on sectors with strong increasing returns to scale. In contrast, when industrial policies are employed, tariffs are mainly used to manipulate terms-of-trade and thereby decrease with trade elasticities.
		
	Third, we find that Nash tariffs under trade war is much higher than those under dual competition. The average Nash tariff under a global trade war is $42\%$, whereas that under global dual competition is $35\%$. Intuitively, if a country is able to efficiently subsidize its high-tech industries with strong economies of scale, it has lower incentives to impose high import tariffs on these industries. Consequently, compared with the global trade war, global dual competition via the combination of tariffs and industrial policies leads to substantially larger welfare gains (or lower welfare losses).
	
	Fourth, we extend our framework to consider global cooperative policies decided by a global social planner that aims to maximize average real income across countries. We find that cooperative tariffs are close to zero, whereas cooperative industrial subsidies are strongly increasing with scale elasticities. Moreover, compared with global cooperation on tariffs, global cooperation on tariffs and industrial subsidies results in larger welfare gains for all major economies except India.
	
	\emph{Related Literature. } This paper is the first attempt to use deep-learning algorithms and machine-learning implementation to solve for optimal policies in general equilibrium models. Our computation framework outperforms previous attempts, \emph{e.g.} \citet{Ossa2014}, based on \citet{Judd2012} in terms of efficiency. This framework is widely applicable for evaluating optimal policies and policy competition in quantitative trade and spatial models.
	
	This paper also relates to recent advances in employing deep learning methods to solve high-dimensional macroeconomic models. \citet{Han2021} and \citet{Sun2023} propose deep learning methods to solve for heterogeneous-agent models with aggregate shocks. \citet{Feng2023} explores optimal taxation with incomplete market. The central objective of this literature is to approximate policy functions with high-dimensional state space using deep learning techniques, and the dimension of action spaces is usually not high. In contrast, our \emph{DL-opt} framework aims at identifying the optimal points in a high-dimensional action space, with the objection functions derived from high-dimensional equilibrium systems. Although our model is static, the dimension of the action space is much higher than those typically addressed in dynamic macro models.
	
	This paper contributes to recent quantitative analysis on trade and industrial policies. \citet{JMWZ2023} characterize optimal trade and industrial policies in the U.S. and China that are uniform across certain industries. This paper, in contrast, computes the fully optimal policies. Additionally, \citet{Lashkaripour2019} and \citet{Bartelme2019} utilize a sufficient statistic approach to characterize optimal trade and industrial policies. However, their reliance on simplification assumptions results in constrained optimal policies, as previously discussed.
	
	The remainder of the paper is organized as follows. In Section \ref{sec:problem}, we propose the general problem and solution framework of \emph{DL-opt}. We then apply this framework to solve optimal trade and industrial policies in a multi-country-multi-sector trade model. In Section \ref{sec:model}, we set the model, define the equilibrium, calibrate the model, and compute the optimal policies using our \emph{DL-opt} framework. Section \ref{sec:policy} characterizes the features of Nash tariffs and industrial subsidies solved in Section \ref{sec:model}. We conclude in Section \ref{sec:Conclusion}.

	\section{DL-opt: General Framework}
	\label{sec:problem}
	
	Our \emph{DL-opt} framework consists of three parts. In this section, we first introduce (i) the nested fixed point (NFXP) algorithm and (ii) automatic implicit differentiation to solve for unilaterally optimal policies. We then discuss (iii) best-response dynamics to solve for Nash policies.
	
	\subsection{Unilaterally Optimal Policies}
	\label{sec:gradient descent}
	
	We start from one granular player, called the government, that decides an action vector $\mathbf{a}$. All atomic players, including households and firms, take $a$ as given. The behaviors of all atomic players and the granular player determine the equilibrium outcome vector $\mathbf{X}$. The general equilibrium is characterized by the following nonlinear system:
	\begin{equation}
		\label{eq:eqm}
		\mathbf{X} = G\left(\mathbf{X},\mathbf{a}\right).
	\end{equation}
	
	The government aims to maximize $W\left(\mathbf{X},\mathbf{a}\right)\in\mathbb{R}$ that depends on both the action $\mathbf{a}$ and the equilibrium outcome $\mathbf{X}$. We can express the government's problem in an MPEC framework:
	\begin{equation}
		\label{eq:MPEC}
		\begin{aligned}
			&\max_{\left(\mathbf{a},\mathbf{X}\right)}W\left(\mathbf{X},\mathbf{a}\right)\\
			\text{s.t. } &\mathbf{X} = G\left(\mathbf{X},\mathbf{a}\right)
		\end{aligned}
	\end{equation}
	However, in many applications $\mathbf{X}$ has much higher dimensions than $\mathbf{a}$. Therefore, the MPEC framework in Equation \eqref{eq:MPEC} incurs the curse of dimensionality and is computationally challenging.
	
	\textbf{Nested Fixed Point (NFPX) Algorithm. } Suppose that we have calculated $\mathbf{X}$ by solving the fixed point in Equation \eqref{eq:eqm} and obtain $\mathbf{X}=\mathbf{X}(\mathbf{a})$ for any given $\mathbf{a}$. Then the government's problem \eqref{eq:MPEC} can be re-expressed as
	\begin{equation}
		\label{eq:NFXP}
		\max_{\mathbf{a}}W\left(\mathbf{X}(\mathbf{a}),\mathbf{a}\right).
	\end{equation}
	This is the nested fixed point (NFXP) algorithm developed by \citet{Rust2000}. It consists of two tiers of loops. 
	\begin{enumerate}
		\item Inner loop: Given $\mathbf{a}$, solve the fixed point $\mathbf{X}$ in Equation \eqref{eq:eqm}.
		\item Outer loop: Update $\mathbf{a}$ by the following general rule:
		\begin{equation}
			\label{eq:GD_discrete}
			\mathbf{a}_{t+1} = \mathbf{a}_t + \gamma \omega\left(\nabla_{\mathbf{a}} W\left(\mathbf{X}(\mathbf{a}_t),\mathbf{a}_t\right)\right),
		\end{equation}
		where $\nabla_{\mathbf{a}} W$ is the gradient of the objective function with respect to $\mathbf{a}$, $\gamma$ is the learning rate, and $\omega(.)$ is a function of the gradient specified by the detailed algorithms.
	\end{enumerate}
		
	\textbf{Fixed Point Iterations. } Given policy $\mathbf{a}$, we calculate $\mathbf{X}$ by the following contraction iterations:
	\begin{equation}
		\label{eq:contraction}
		\mathbf{X}_{t+1} = G\left(\mathbf{X}_{t},\mathbf{a}\right).
	\end{equation}
	
	\citet{Rust2000} argues that the calculation of fixed point $\mathbf{X}$ can be accelerated by utilizing Newton-Kantorovich iterations after contraction iterations. The Newton-Kantorovich iterations have the following form:
	\begin{equation}
		\mathbf{X}_{t+1} = \mathbf{X}_{t} - \left[\mathbf{I}-\nabla_{\mathbf{X}}G\left(\mathbf{X}_{t},\mathbf{a}\right)\right]^{-1}\left[\mathbf{X}_t-G\left(\mathbf{X}_{t},\mathbf{a}\right)\right],
	\end{equation}
	where $\mathbf{I}$ is the identity matrix and $\nabla_{\mathbf{X}}G$ is the gradient of equilibrium system with respect to $\mathbf{X}$. However, this method is computationally burdensome when $G$ and $\mathbf{X}$ are high-dimensional since it requires computing the gradient $\nabla_{\mathbf{X}}G$ for each $\mathbf{X}_t$. In the \emph{DL-opt} framework, we prefer using contraction iterations in Equation \eqref{eq:contraction} to obtain the fixed point $\mathbf{X}_T$.
		
	\textbf{Automatic Implicit Differentiation. } The key challenge to solve the problem in Equation \eqref{eq:NFXP} is to compute $\nabla_{\mathbf{a}}W\left(\mathbf{X}(\mathbf{a}),\mathbf{a}\right)$ in Equation \eqref{eq:GD_discrete}. By the chain rule, it can be expressed as
	\begin{equation}
		\label{eq:gradient}
		\nabla_{\mathbf{a}} W\left(\mathbf{X}_T,\mathbf{a}\right)=\left[\nabla_{\mathbf{a}}W(\mathbf{X}_T,\mathbf{a})+\nabla_{\mathbf{X}}W(\mathbf{X}_T,\mathbf{a})\nabla_{\mathbf{a}}\mathbf{X}_T\right],
	\end{equation}
	where $\mathbf{X}_T$ is the fixed point in Equation \eqref{eq:eqm} under $\mathbf{a}$.
	
	Due to high-dimensionality of both $\mathbf{X}$ and $\mathbf{a}$, it is generally difficult to compute $\nabla_{\mathbf{a}}\mathbf{X}_T$. However, by the implicit function theorem, we have
	\begin{equation}
		\label{eq:implicit}
		\nabla_{\mathbf{a}}\mathbf{X}_T=-\left[\nabla_{\mathbf{X}}G\left(\mathbf{X}_T,\mathbf{a}\right)-\mathbf{I}\right]^{-1}\nabla_{\mathbf{a}}G\left(\mathbf{X}_T,\mathbf{a}\right).
	\end{equation}
		
	We utilize automatic differentiation in deep learning to efficiently calculate $\nabla_{\mathbf{X}}G\left(\mathbf{X}_T,\mathbf{a}\right)$ and $\nabla_{\mathbf{a}}G\left(\mathbf{X}_T,\mathbf{a}\right)$ at the point $\mathbf{X}_T$. Equation \eqref{eq:implicit} combines the implicit function theorem with automatic differentiation. Therefore, it is called \emph{automatic implicit differentiation}.\footnote{The idea of automatic implicit differentiation is first proposed in Jeffrey Sun's computation notes. See \url{https://jeffreyesun.com/auto_implicit_diff.pdf}.}

    Now we briefly introduce the idea of automatic differentiation. Automatic differentiation is particularly efficient when $G(.,.)$ is high-dimensional and consists of many elementary operations and functions. Suppose that $\mathbf{X}$ has $K$ dimensions. Then to calculate $\nabla_{\mathbf{X}}G\left(\mathbf{X}_T,\mathbf{a}\right)$ using finite difference, we need to evaluate $G(.,.)$ for $K+1$ times. This can be very challenging when $K$ is large. 
    
    Automatic differentiation, in contrast, decomposes $G(.,.)$ into many intermediate variables. Each intermediate variable can be computed by either (i) the elements of $\mathbf{X}$, or (ii) other intermediate variables through elementary operations and functions. These intermediate variables formulate a computation graph that connects $G(.,.)$ with $\mathbf{X}$. 
    
    The key advantage of automatic differentiation is that many intermediate variables are common in calculating $\frac{\partial G}{\partial X_i}$ and $\frac{\partial G}{\partial X_j}$ for $i\neq j$. Therefore, we do not need to evaluate the high-dimensional $G(.,.)$ for $K$ times to compute $\nabla_{\mathbf{X}}G\left(\mathbf{X}_T,\mathbf{a}\right)$. In Appendix \ref{appsec:AD}, we provide a toy example to illustrate the advantage of automatic differentiation in detail.
    
\begin{figure}[htbp]
    	\centering
    	\subfloat[Policy Optimization]{\includegraphics[width=0.5\textwidth]{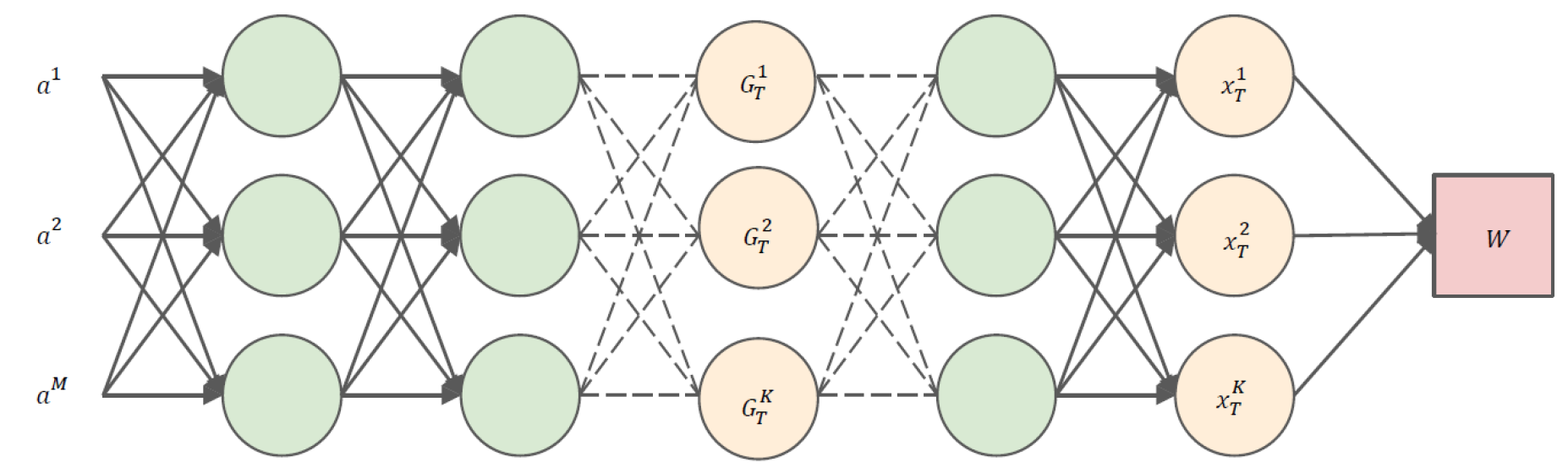}}\\
    	\subfloat[Neural Networks]{\includegraphics[width=0.5\textwidth]{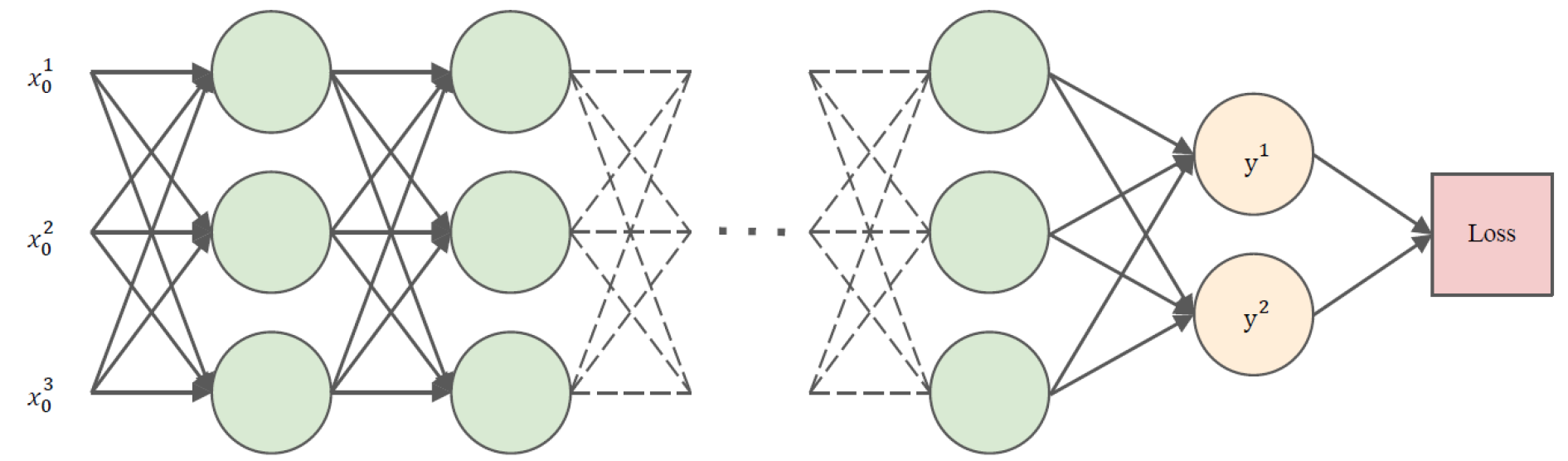}}
    	\caption{Analogy between Policy Optimization and Neural Networks}
    	\label{fig:neural_network_analogy}
    \end{figure}

    Repeated application of automatic differentiation to Equation \eqref{eq:implicit} and \eqref{eq:gradient} establishes a connection between $W(.,.)$ and $\mathbf{a}$ through multiple layers of intermediate variables and elementary operations. Figure \ref{fig:neural_network_analogy} draws a parallel between policy optimization and neural networks. In this analogy, policies $\mathbf{a}$ are like neural network parameters, intermediate variables correspond to nodes in hidden layers, and the objective function $W$ is analogous to the loss function. Policy optimization adjusts policies $\mathbf{a}$ to maximize $W$ via layers of intermediate variables (Panel (a) of Figure \ref{fig:neural_network_analogy}), similar to how neural networks adjust parameters to approximate outputs and minimize the loss function (Panel (b) of Figure \ref{fig:neural_network_analogy}). Consequently, gradient computation through automatic implicit differentiation mirrors the forward-backward propagation in neural networks depicted in Panel (b). This powerful analogy suggests that well-developed deep learning tools can effectively implement automatic differentiation for any well-defined function $G$.
       
	\subsection{Nash Policies}
	
	We proceed by considering multiple granular players ($i=1,\ldots,N$) and Nash policies. In particular, play $i$ decides action $\mathbf{a}_i$, taking others' actions $\mathbf{a}_{-i}$ as given. The best response function of player $i$ can be given by
	\begin{equation}
		\label{eq:BR}
		b^*_i\left(\mathbf{a}_{-i}\right)\equiv \arg\max_{\mathbf{a}_i}W\left(\mathbf{X}(\mathbf{a}_i,\mathbf{a}_{-i}),\mathbf{a}_i,\mathbf{a}_{-i}\right),
	\end{equation}
	where $\mathbf{X}$ is the fixed point in $\mathbf{X} = G\left(\mathbf{X},\mathbf{a}_i, \mathbf{a}_{-i}\right)$. Given $\mathbf{a}_{-i}$, $b^*_i\left(\mathbf{a}_{-i}\right)$ can be computed using the NFXP algorithm and automatic implicit differentiation in our \emph{DL-opt} framework described in Section \ref{sec:gradient descent}.
	
	With the best-response function, $b^*_i\left(\mathbf{a}_{-i}\right)$, in hand, we solve for Nash equilibria, $\left(\mathbf{a}^*_i\right)_{i=1}^{N}$, where $\mathbf{a}^*_i\in b^*_i\left(\mathbf{a}_{-i}\right)$ for $i=1,\ldots,N$. We transform the static problem into a system of best-response dynamics:
	\begin{equation}
		\label{eq:BR_discrete}
		\mathbf{a}_i^{t+1} = \eta_t b^*_i\left(\mathbf{a}_{-i}^t\right) +(1-\eta_t) \mathbf{a}_i^t,\quad \forall i=1,2,\ldots,N,
	\end{equation}
	where $\eta_t\in (0,1]$ is the stepsize. 
	
	The best-response dynamics capture the idea of players adjusting their strategies iteratively in response to the strategies chosen by others. In the iterative algorithm expressed by Equation \eqref{eq:BR_discrete}, players subsequently choose their best responses, given others' strategies. The key to this algorithm is the sequence of players' actions. \cite{HJMP+2021} consider two specific playing sequences for determining the order in which players update their actions during the game. One approach is a fixed cyclic order, where players take turns adjusting their strategies in a predetermined order. The other approach is the random playing sequence, where, at each time step, a player is selected uniformly at random from among all players to update their strategy. 
	
	Our \emph{DL-opt} framework utilizes a random shuffle playing sequence that combines the features of the two sequences above. This sequence introduces randomness into the playing order, which helps to reduce positional advantages that may arise in a clockwork playing sequence. Shuffling the playing order randomly at the beginning of each round ensures that each player has an equal opportunity to adjust their actions in each round. The best-response dynamics with random shuffle playing sequence is described in Algorithm \ref{alg:brd}.
	
	\begin{center}
		\begin{algorithm}[htbp]\footnotesize
			\begin{enumerate}
				\item Guess the initial action profile of all players $\mathbf{a}^{0}$.
				\item Repeat until converge: 
				\begin{itemize}
					\item Random shuffle playing sequence for each round
					\item For the $i$-th player $s(i)$ in the shuffled sequence $s$
					\begin{enumerate}
						\item Compute the best response of the player $s(i)$ in sequence $s$
						\begin{equation}
							\mathbf{a}_{s(i)}^{t+1} = \eta_t b^*_{s(i)}\left(\mathbf{a}_{-s(i)}^t\right) +(1-\eta_t) \mathbf{a}_{s(i)}^t.
						\end{equation}
						\item Update $\mathbf{a}_{-n}^{t+1}=\mathbf{a}_{-n}^{t}$ for all $n\neq s(i)$.
					\end{enumerate}
				\end{itemize}
			\end{enumerate}
			\caption{Best-response dynamics with random shuffle playing sequence}
			\label{alg:brd}
		\end{algorithm}
	\end{center}
	
	\begin{figure}
		\centering
		\includegraphics[width=0.5\textwidth]{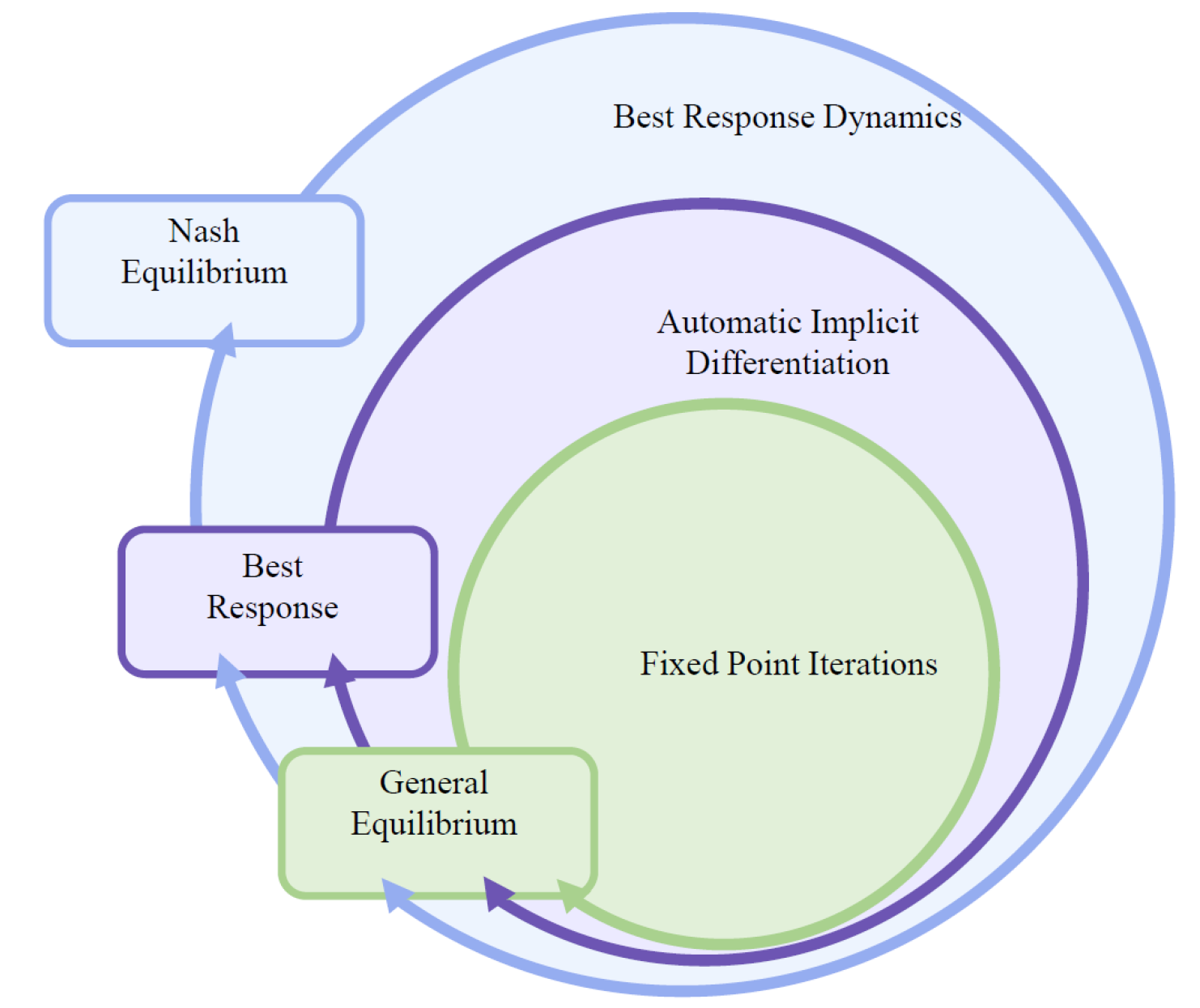}
		\caption{Calculation Flows of the DL-opt Framework}
		\label{fig:flowchart}
	\end{figure}
	
	Figure \ref{fig:flowchart} summarizes the calculation flows of our DL-opt framework. First, the general equilibrium outcome $\mathbf{X}$ is determined using fixed point iteration within the inner loop of the NFXP algorithm. Second, the best response function is computed via gradient descent, supported by automatic implicit differentiation. Finally, the Nash equilibrium is achieved through best response dynamics with a randomly shuffled playing sequence.

	\section{Model, Calibration, and Computation}
	\label{sec:model}
	
	In this section, we present a multi-country-multi-sector general equilibrium trade model with input-output linkages and sectoral scale economies and solve the optimal (unilaterally and mutually) trade and industrial policies in this model. This model, a la \citet{Bartelme2019}, \citet{Lashkaripour2019}, and \citet{JMWZ2023}, provides a laboratory for quantifying global trade and industrial policy competitions. We first introduce the model and its equilibrium, then discuss the model's calibration, and finally utilize our \emph{DL-opt} framework to compute Nash tariffs and industrial subsidies. 
	
	\subsection{Model and Equilibrium}
	
	Consider $N$ countries with labor endowments $\left(L_{i}\right)_{i=1}^{N}$. There are $J$ sectors. Workers cannot relocate between countries but are free to move across sectors. Each sector $j$ consists of a unit mass of varieties. The representative consumer in country $i$ has a Cobb-Douglas preference over $J$ sectors, with share $\alpha_i^j$ for sector $j$, and a CES preference over varieties within in sector, with the elasticity of substitution $\sigma_j>1$.
	
	Each variety is produced under perfect competition using labor and composite intermediates. The unit cost of variety $\omega$ of sector $j$ in country $i$ is $c_{i}^{j}(\omega)=\frac{c_{i}^{j}}{z_{i}^{j}(\omega)}$,
	where $c_{i}^{j}$ is defined as 
	\begin{equation}
		c_{i}^{j}=\frac{1}{\left(L_{i}^{j}\right)^{\psi_{j}}}w_{i}^{\beta_{i}^{j}}\left[\prod_{s=1}^{J}(P_{i}^{s})^{\gamma_{i}^{sj}}\right]^{1-\beta_{i}^{j}},
	\end{equation}
	where $w_i$ is the wage in country $i$ and $\beta_{i}^{j}$ represents the value-added share, $P_{i}^{s}$ is the price index of good $s$ in country $i$ and $\gamma_{i}^{sj}$ represents the share of good $s$ in producing good $j$ in sector $j$ in country $i$, and $L_{i}^{j}$ is the labor allocated to sector $j$ of country $i$ and $\psi_{j}\geq 0$ characterizes the external economies of scale in sector $j$.	
	
	Following \citet{Eaton2002}, we assume that $z_{i}^{j}(\omega)$ is drawn independently from a Frech\'et distribution:
	\begin{equation}
		\mathbb{P}\left(z_{i}^{j}(\omega)\leq z\right)=\exp\left(-T_{i}^{j}z^{-\theta_{j}}\right)
	\end{equation}
	where $z>0$ and $\theta_{j}>\max\{\sigma_{j}-1,1\}$.
	
	Exporting good $j$ from country $i$ to $n$ incurs an iceberg trade cost $\tau_{in}^{j}\geq 1$. Moreover, we consider two policies. First, an ad valorem tariff $t_{in}^{j}\geq 0$ is imposed by importing country $n$ on goods $j$ imported from country $i$, with $t_{ii}^{j}=0$. Second, an ad valorem output tax $s_{i}^{j}\in[0,1)$ is imposed by production country $i$ on goods $j$.
	
	Based on the property of Frech\'et distribution and the ideal price index of CES preferences, the sectoral price index can be expressed as 
	\begin{equation}
		\label{eq:price}
		P_{n}^{j}=\left(\sum_{i=1}^{N}T_{i}^{j}\left[c_{i}^{j}\tau_{in}^{j}\left(1+t_{in}^{j}\right)\left(1-s_i^j\right)\right]^{-\theta_{j}}\right)^{-\frac{1}{\theta_{j}}}.
	\end{equation}
	
	The expenditure share of country $n$ on good $j$ from country $i$ is given by 
	\begin{equation}
		\pi_{in}^{j}=\frac{X_{in}^{j}}{X_{n}^{j}}=\frac{T_{i}^{j}\left[c_{i}^{j}\tau_{in}^{j}\left(1+t_{in}^{j}\right)\left(1-s_i^j\right)\right]^{-\theta_{j}}}{\left(P_{n}^{j}\right)^{-\theta_{j}}},
	\end{equation}
	where $X_{in}^{j}$ is the expenditure of country $n$ on good $j$ from country $i$ and $X_n^j$ is the total expenditure on good $j$ in country $n$.
	
	Sectoral employment satisfies 
	\begin{equation}
		w_{i}L_{i}^{j}=\beta_{i}^{j}\sum_{n=1}^{N}\frac{X_{in}^{j}}{(1+t_{in}^{j})(1+e_{in}^{j})}.\label{eq:labor}
	\end{equation}
	
	The wage is determined by labor market clearing 
	\begin{equation}
		\sum_{j=1}^{J}L_{i}^{j}=L_{i}.\label{eq:labor_clear}
	\end{equation}
	
	Assuming that export tariffs (if any) are collected before import tariffs and all tax revenues or subsidy expenditures are transferred to workers as lump-sum payments, we can express the total final income as 
	\begin{equation}
		Y_{i}=w_{i}L_{i}+\sum_{j=1}^{J}\sum_{n=1}^{N}\frac{e_{in}^{j}}{1+e_{in}^{j}}X_{in}^{j}+\sum_{j=1}^{J}\sum_{k=1}^{N}\frac{t_{ki}^{j}}{(1+t_{ki}^{j})(1+e_{ki}^{j})}X_{ki}^{j}.
	\end{equation}
	
	The sectoral expenditure can be expressed by
	\begin{equation}
		X_{i}^{j}=\alpha_{i}^{j}Y_{i}+\sum_{s=1}^{J}(1-\beta_{i}^{s})\gamma_{i}^{js}\sum_{n=1}^{N}\frac{X_{in}^{s}}{(1+t_{in}^{s})(1+e_{in}^{s})}.\label{eq:expenditure}
	\end{equation}
	
	\begin{definition}[Equilibrium]
		\label{def:equilibrium}
		Given parameters $\left\{\theta_{j},\psi_{j},\alpha_{i}^{j},\beta_{i}^{j},\gamma_{i}^{sj},L_{i},s_i^j,t_{in}^{j},T_{i}^{j},\tau_{in}^{j}\right\}$,
		the equilibrium consists of $\left(w_{i},L_{i}^{j},P_{i}^{j},X_{i}^{j}\right)$ such that (i) $\left(L_i^j\right)$ satisfy Equation \eqref{eq:labor}; (ii) $\left(w_i\right)$ are determined by Equation \eqref{eq:labor_clear}; (iii) $\left(P_i^j\right)$ satisfy Equation \eqref{eq:price}; (iv) $\left(X_i^j\right)$ satisfy Equation \eqref{eq:expenditure}.
	\end{definition}
	
	The welfare in country $n$ can be measured by its real income, $W_n\equiv\frac{Y_n}{P_n}$, where the aggregate price index for final consumption goods can be expressed as $P_{n}=\prod_{j=1}^{J}\left(P_{n}^{j}\right)^{\alpha_{n}^{j}}$.
		
	The equilibrium system in Definition \ref{def:equilibrium} consists of $3NJ+N$ nonlinear equations that relate to $3NJ+N$ unknown variables, which can be solved when given a numeraire. However, this system poses a challenge as it depends on a complex set of parameters, including $T_{i}^{j}$, $\tau_{in}^{j}$, which are difficult to calibrate. To address this issue, we use the ``exact-hat" algebra developed by \cite{DEK2008} to compute changes in equilibrium outcomes relative to changes in exogenous shocks. We denote the value of any variable after the change as $Z'$ and use the notation $\widehat{Z}=Z'/Z$. Given the values of $\left(\alpha_{i}^{j}, \beta_{i}^{j}, \gamma_{i}^{sj}; \psi_{j},\theta_{j}\right)$, as well as data on $\left(X_{in}^{j},t_{in}^{j},s_i^j\right)$, we can compute the changes in equilibrium outcomes, $\left(\widehat{w}_{i}^{j}, \widehat{L}_{i}^{j}, \widehat{P}_{i}^{j},\widehat{X}_{i}^{j}\right)$, by solving a system of $3NJ+N$ nonlinear equations. In our application, this system is the equilibrium system $G(.,.)$ described in Section \ref{sec:gradient descent}. The details of this system are presented in Appendix \ref{sec:exact-hat}. 
	
	The optimization problem faced by each country $i$ involves maximizing the change in its welfare, $\hat{W}_i$, by choosing $\left(t_{ki}^j,s_i^j\right)$ for all $k\neq i$ and $j$. The details of this problem are also presented in Appendix \ref{sec:exact-hat}.
	
	\subsection{Calibration}
	
	The ``exact-hat" algebra requires data on bilateral trade shares $\left(\pi^j_{in}\right)$, sectoral consumption shares $\left(\alpha^j_i\right)$, sectoral value-added shares $\left(\beta^j_i\right)$, sectoral expenditure $\left(X^j_n\right)$, input expenditure shares $\gamma^{js}_i$, and initial policies $\left(t^j_{in}, s_i^j\right)$. We also need the values of parameters $\left(\psi_j, \theta_j\right)$.
	
	We consider 6 major economies (US, China, Japan, EU, Brazil, and India) and the rest of the world (ROW).\footnote{European Union (EU) includes 28 countries, including the UK.} We utilize the OECD Inter-Country Input-Output Database (ICIO) for 2017 to extract internationally comparable data on country-sectoral production, value-added, bilateral trade flows, and input-output linkages. The ICIO table includes 22 tradable sectors and 22 nontradables.\footnote{The ICIO has 45 industries. We disregard the last one, which is ``Activities of households as employers; undifferentiated goods- and services-producing activities of households for own use'' because it has many zeros. See OECD. (2021) OECD Inter-Country Input-Output Database, http://oe.cd/icio.} We get the initial import tariffs from the World Integrated Trade System (WITS) for 2017 and assume that initially $s_i^j = 0$ for all $(i,j)$.
	
	We calibrate $\left(\psi_j,\theta_j\right)$ from \citet{Lashkaripour2019}. The calibrated values of $\left(\psi_j,\theta_j\right)$ in Appendix \ref{appsec:calibration}. \citet{Lashkaripour2019} recover $\psi_j$ from the effects of variation in sector size on equilibrium quantities, exploiting variation in countries' population and preferences as instruments.

	\subsection{Computation}
	\label{sec:Computation}
	
	We utilize our \emph{DL-opt} framework to compute $\left(t_{ki}^j,s_i^j\right)$ that maximize $\hat{W}_i$ for each country $i$. We select the Adaptive Moment Estimation (ADAM) algorithm to implement gradient descent in Equation \eqref{eq:GD_discrete}. The procedures and advantages of the ADAM algorithm are presented in Appendix \ref{appsec:adam}.
		
	Our stochastic best-response dynamics is implemented using the PyTorch machine learning open-source framework.\footnote{See \cite{PGM+2019} for the introduction of PyTorch.} PyTorch allows us to take advantage of automatic differentiation, enabling us to efficiently compute gradients through forward and backward propagation. This feature is crucial for obtaining gradients with respect to the action in our problem effectively. Additionally, PyTorch offers support for various machine learning optimization methods, including SGD, ADAM, Adagrad, and RMSprop. This versatility allows us to choose the most appropriate gradient-based learning method for our specific task. 
	
	Due to the capabilities of the PyTorch framework, our stochastic best-response dynamic program is easily implemented. The combination of PyTorch's
	flexibility and efficient optimization methods contributes to the rapid convergence of our stochastic best-response dynamics. It's worth mentioning that other machine learning frameworks such as TensorFlow or Google JAX\footnote{See \cite{AAB+2016} and \cite{FJL2018} for the details of these frameworks.} could also be utilized for similar purposes.
		
	Our baseline exercise involves computing Nash tariffs and industrial subsidies for all 7 economies in our calibrated model (US, China, Japan, EU, Brazil, India, and the rest of the world). We refer to this baseline scenario as \emph{global dual competition}. The computation statistics of our framework in this baseline scenario are summarized in Table \ref{table:hyperparameters}. It takes about 5 hours in a single-core personal computer to compute Nash tariffs and industrial policies for all 7 economies and 44 industries.
	
		\begin{table}[htbp]\footnotesize
        \centering
        \caption{Computation statistics of Solving for Global Dual Competition}
        \label{table:hyperparameters}
        
				\begin{tabular}{cc}
					\toprule 
					\textbf{Information} & \textbf{Value}\\
					\midrule
					ML Framework & PyTorch\\
					Device & CPU\\
					Num. epochs & $\sim20$\\
					Iterations per player/epoch & $\sim50$\\
					Num. players & 7\\
					Playing sequence & Random Shuffle\\
					Optimizer & ADAM\\
					Anneal learning rate & False\\
					Learning rate & $10^{-4}\sim10^{-3}$\\
					Is clipping grad & \{True, False\}\\
					Max grad norm & 10.0\\
                    Max Computation Time & 5h \\
					\bottomrule
				\end{tabular}
		\end{table}
	
	We compared the computational efficiency of our \emph{DL-opt} framework with widely-used nonlinear solvers such as Knitro. The \emph{DL-opt} framework solves for unilaterally optimal industrial subsidies in China in less than \textbf{one hour}, whereas Knitro, using interior-point methods without analytical Jacobian and Hessian matrices, takes \textbf{seven hours} to achieve the same solution. For this small-scale problem, our \emph{DL-opt} framework is approximately ten times more efficient than Knitro under these conditions.	
	
\begin{figure}
		\centering
		\includegraphics[width=0.75\textwidth]{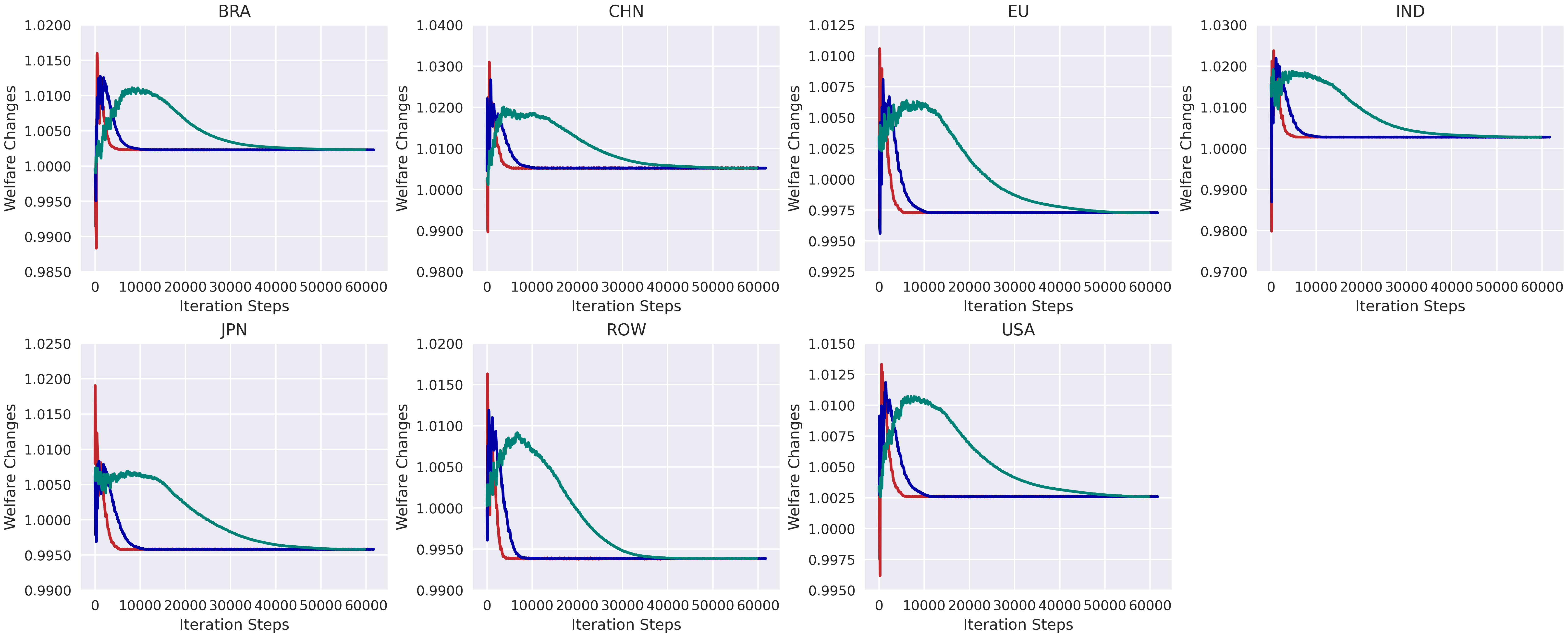}
		
		\caption{Iteration Curve for Nash Equilibrium of Global Dual Competition}
		\label{fig:nash_iteration_curve}
		
		\vspace{0.2cm}
		\footnotesize
		(Note: The three lines represent distinct computational experiments. The red line corresponds to a learning rate of 0.001, the blue line to a learning rate of 0.0005, and the green line to a learning rate of 0.0001.)
	\end{figure}

    Figure \ref{fig:nash_iteration_curve} illustrates the convergence path of our algorithm in each country. The three lines denote separate computational experiments, each conducted under different learning rates. Despite these variations, the outcomes across all experiments remain consistent. It also shows that even under relatively strong scale economies, our algorithm converges steadily after about 10,000 iterations. 

\begin{figure}
    	\centering
    	\includegraphics[width=0.75\textwidth]{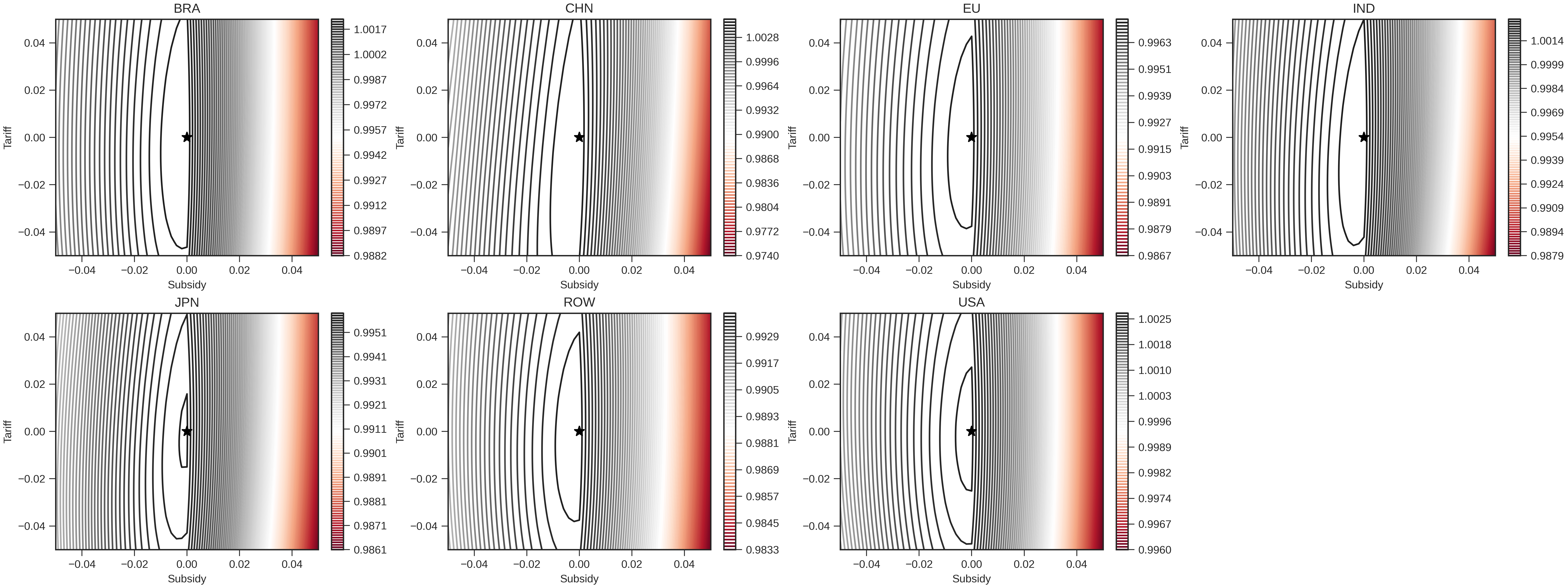}
    	
    	\caption{Landscape Near Nash Equilibrium for Global Dual Policy Competition	with Scale Economies}
    	\label{fig:nash_landscape_scale}
    \end{figure}

    Finally, we examine the optimality of our solutions. Figure \ref{fig:nash_landscape_scale} illustrates a landscape near Nash equilibrium for global dual policy competition, suggesting the local optimality of our solutions. Moreover, welfare decreases dramatically when subsidy rates exceed their optimal levels but decreases slowly when they are below the optimal. This is mainly because excessive subsidies require additional subsidy expenses and thereby lead to greater welfare losses.

	\section{Optimal Trade and Industrial Policies}
	\label{sec:policy}
	
	Our \emph{DL-opt} framework is efficient in solving for high-dimensional continuous optimal policies in multi-country-multi-sector general equilibrium models. Compared with the existing methods, our framework takes into account (i) rich heterogeneity across countries and sectors and (ii) combinations of policy instruments. In this section, we focus on global dual competition in which each country decides its import tariffs and sector-specific production subsidies to maximize its real income, given other countries' tariffs and industrial subsidies. To understand the interdependence of trade and industrial policies, we further compare tariffs under global dual competition with those under the global trade war. 
	
	We discuss our quantitative results as follows. First, we depict the sectoral heterogeneity of non-cooperative tariffs and industrial subsidies solved using our \emph{DL-opt} framework. Second, we explore how non-cooperative tariffs will change if countries are allowed to compete via industrial subsidies. Third, we extend our framework to solve global cooperative tariffs and industrial subsidies. Finally, we show that in global competition, industrial subsidies have to be correctly specified to generate welfare gains.
	
	\subsection{Sectoral Heterogeneity of Optimal Policies}
	
	We use our \emph{DL-opt} framework to solve for (unilaterally and mutually) optimal tariffs and industrial subsidies, taking into account rich heterogeneity across countries and sectors. \citet{Lashkaripour2019} has provided analytical characterizations of unilaterally optimal tariffs and industrial policies. However, their analytical results are built on the ``internal cooperation" assumption: the relative wages among other countries remain unchanged under the optimal policies. Utilizing our \emph{DL-opt} framework, we investigate changes in relative wages under China's fully optimal policies to see whether this assumption is restrictive in solving for the fully optimal policies. The result suggests that the ``internal cooperation" assumption is restrictive, and, therefore, the analytical unilaterally optimal tariffs and industrial policies in \citet{Lashkaripour2019} are constrained optimal. The detailed results are presented in Appendix \ref{appsec:internal_cooperation}.
		
\begin{figure}[htbp]
		\centering
		\subfloat[Tariffs (trade war)]{\includegraphics[width=0.33\textwidth]{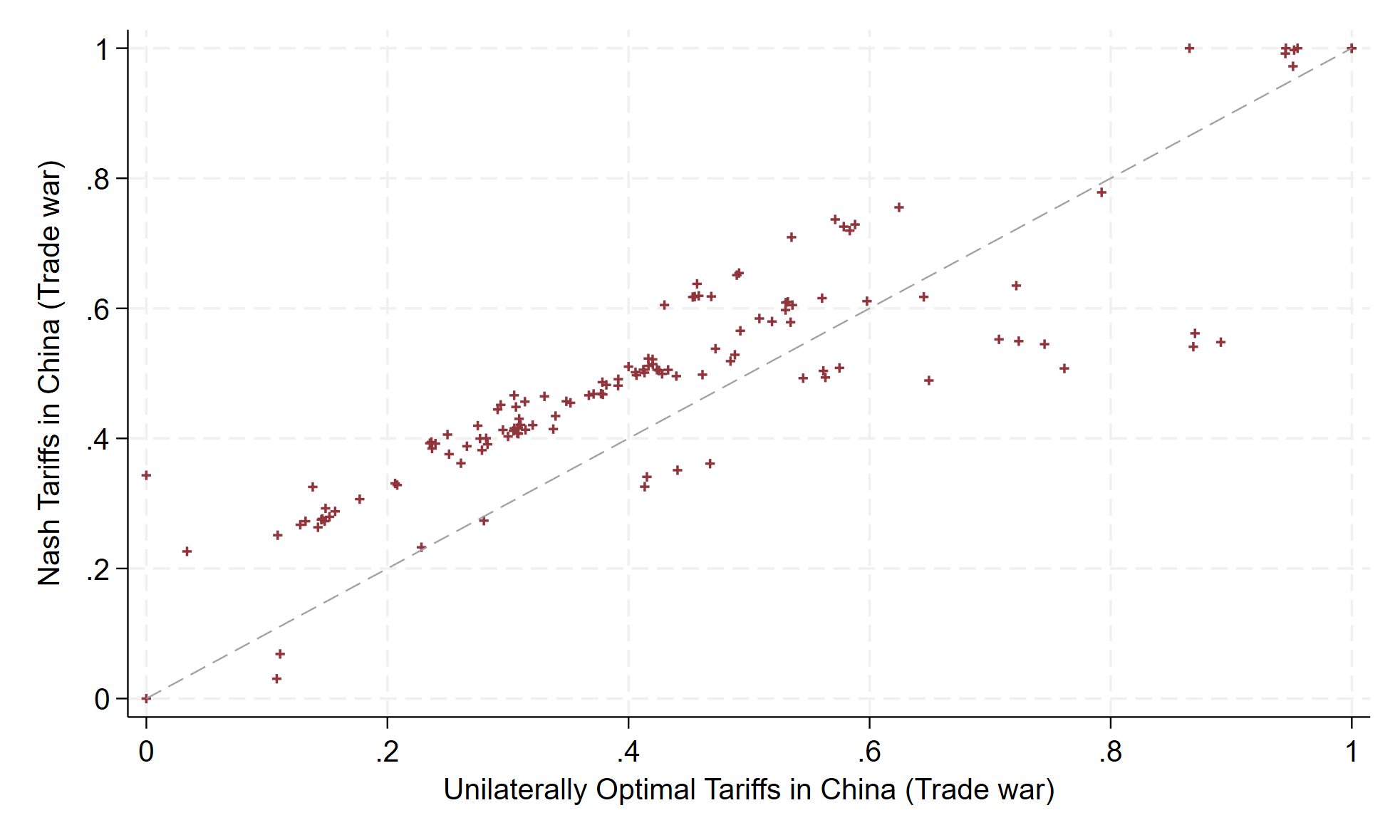}}
		\subfloat[Tariffs (dual)]{\includegraphics[width=0.33\textwidth]{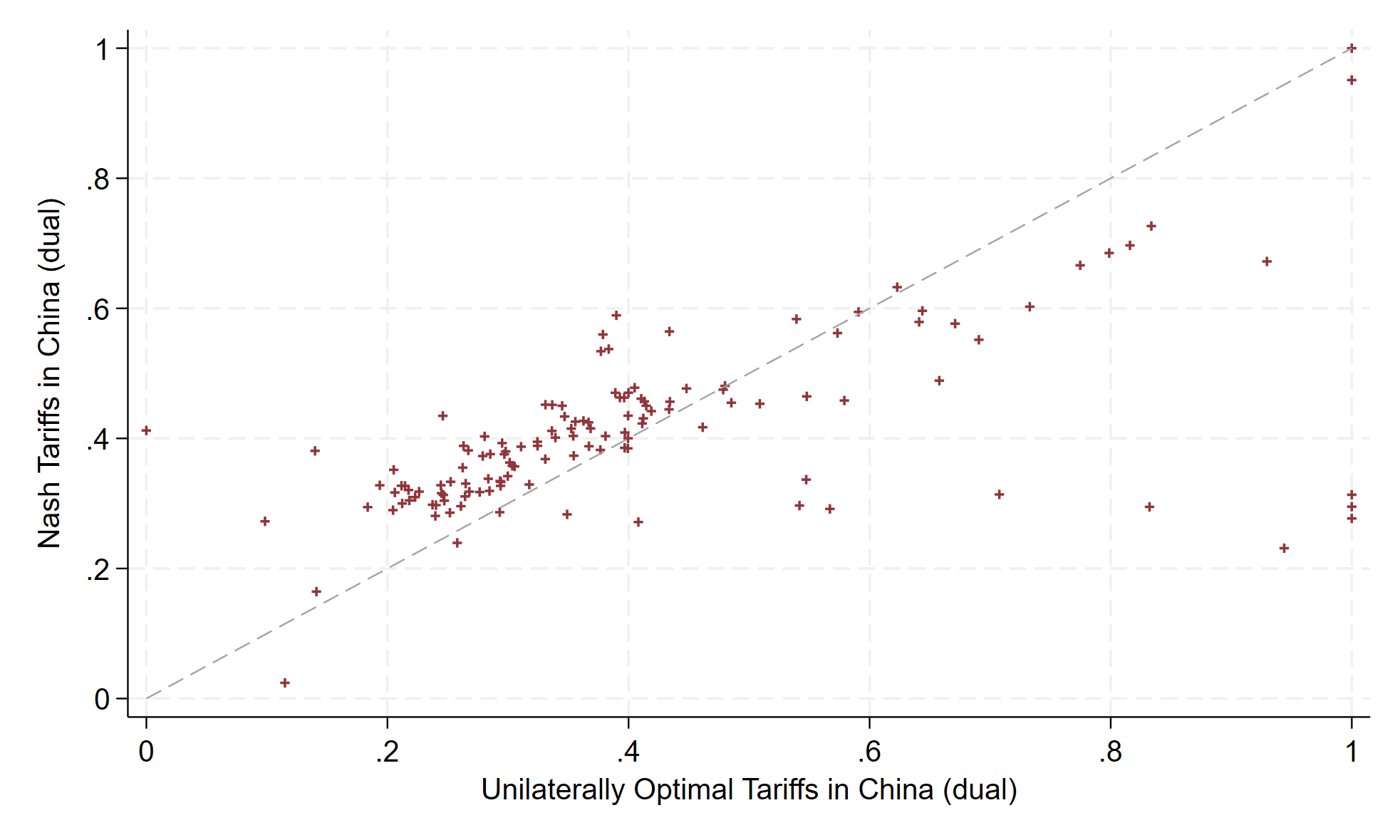}}
		\subfloat[Subsidies (dual)]{\includegraphics[width=0.33\textwidth]{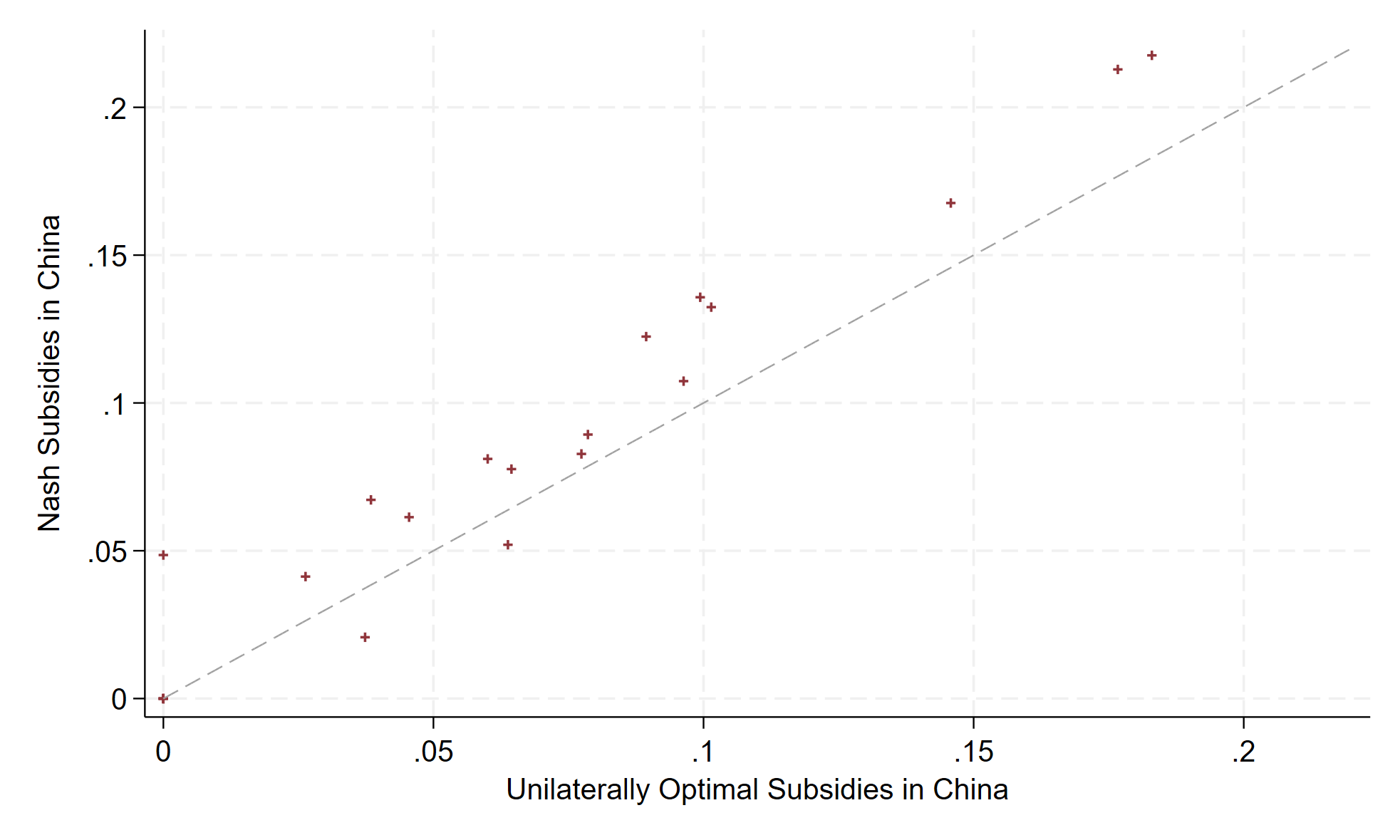}}
		
		\caption{Unilaterally Optimal vs. Nash Policies in China}
		\label{fig:uni_nash_compare}
		
		\vspace{0.2cm}
		\scriptsize
		(Notes: The dash line is the 45-degree line.)
	\end{figure}

	We proceed by depicting optimal policies computed using our \emph{DL-opt} framework. Figure \ref{fig:uni_nash_compare} compares the unilaterally optimal and Nash policies in China, suggesting that Nash tariffs and industrial subsidies are considerably higher than unilaterally optimal policies. Intuitively, other countries' optimal trade and industrial policies would shrink China's production scale in increasing-returns-to-scale industries and thereby strengthen China's incentives to promote production scale by tariffs and industrial subsidies. This result highlights the importance of characterizing mutually optimal (Nash) policies in understanding global competition.
	
	We then illustrate how our Nash policies relate to sectoral trade elasticities, $\left(\theta_j\right)$, and scale elasticities $\left(\psi_j\right)$. To this end, we regress $\log\left(t_i^j\right)$ (under trade war and under dual competition) on $\log\left(\theta_j\right)$ and $\log\left(\psi_j\right)$, controlling for importer and exporter fixed effects. Moreover, we regress $\log\left(\frac{1}{1-s_i^j}\right)$ on $\log\left(\theta_j\right)$ and $\log\left(\psi_j\right)$, controlling for country fixed effects. The results are reported in Table \ref{tab:nash_theta_psi}.
	
	\begin{table}[htbp]\footnotesize
		\centering
		\caption{Nash Policies and Trade and Scale Elasticities}
		\label{tab:nash_theta_psi}
		\begin{tabular}{lccc}\toprule
			Dep. Var:  & $\log\left(\frac{1}{1-s_i^j}\right)$-dual & $\log\left(t_{in}^j\right)$-dual   & $\log\left(t_{in}^j\right)$-trade war  \\\cmidrule(lr){2-3}\cmidrule(lr){4-4}
			& (1) & (2) & (3) \\
			$\log\left(\theta_j\right)$ &   $-0.01$  & $-0.15^{***}$ & $0.09^{***}$  \\
			&  $(0.01)$ & $(0.01)$   & $(0.03)$ \\
			$\log\left(\psi_j\right)$   & $0.12^{***}$   &   $0.20^{***}$     &  $0.64^{***}$ \\
			& $(0.01)$  &  $(0.02)$      & $(0.03)$ \\
			Exp. FE &  $\checkmark$     &   $\checkmark$    &  $\checkmark$     \\
			Imp. FE      &      &  $\checkmark$ &  $\checkmark$    \\
			\# obs  & 154 & 924      & 923  \\
			$R^2$   & 0.63 & 0.30    & 0.31 \\
			\bottomrule
			
			\multicolumn{4}{l}{
				\begin{minipage}{10cm}
					\vspace{0.5\baselineskip}
					\scriptsize \textit{Note: ``trade war" refers to cases where players can modify their import tariffs solely. ``dual" refers to cases where players have the flexibility to adjust both their industry subsidies and import tariffs. }       		
			\end{minipage}}
		\end{tabular}
	\end{table}
	
	First, Nash industrial subsidies under dual competition are positively correlated with scale elasticities $\left(\psi_j\right)$ but not correlated with trade elasticities $\left(\theta_j\right)$. As discussed in \citet{Lashkaripour2019}, industrial subsidies are mainly used to address misallocation across sectors: subsidizing sectors with larger $\psi_j$ could lead to greater gains from the increase in production scale, whereas the terms-of-trade effect is not a concern in determining Nash industrial subsidies. Our numerical result about Nash industrial subsidies is consistent with theoretical results in \citet{Lashkaripour2019}.
		
	Second, Nash tariffs (under trade war and under dual competition) are positively correlated with scale elasticities, $\left(\psi_j\right)$. This is due to the home-market effect: the increase in import tariffs would induce more domestic production, the benefit of which is larger in sectors that have stronger increasing returns to scale. Notice that the positive correlation between Nash tariffs and scale elasticities is much stronger under trade wars than that under dual competition (comparing coefficients in Column (2) and (3) in Table \ref{tab:nash_theta_psi}). When countries can compete via industrial policies, they mainly rely on industrial subsidies to address the misallocation of sectoral production led by external economies of scale. In contrast, when they can only manipulate tariffs, tariffs have to be prohibitively high on sectors with strong increasing returns to scale.
	
	Third, Nash tariffs under dual competition are negatively correlated with trade elasticities, $\left(\theta_j\right)$. As figured out by \citet{Eaton2002}, smaller $\theta_j$ indicates stronger within-sector comparative advantage and thereby larger terms-of-trade effects of tariffs. Under dual competition, tariffs are used to address terms-of-trade effects and thereby negatively correlated with $\left(\theta_j\right)$. Notably, Nash tariffs under trade war are slightly positively correlated with $\theta_j$. This is because when industrial subsidies are absent, tariffs have to be used to address misallocation across sectors. In this case, the terms-of-trade effect is not the primary concern in determining Nash tariffs.
	
\begin{figure}[htbp]
		\centering
		\includegraphics[width=0.5\textwidth]{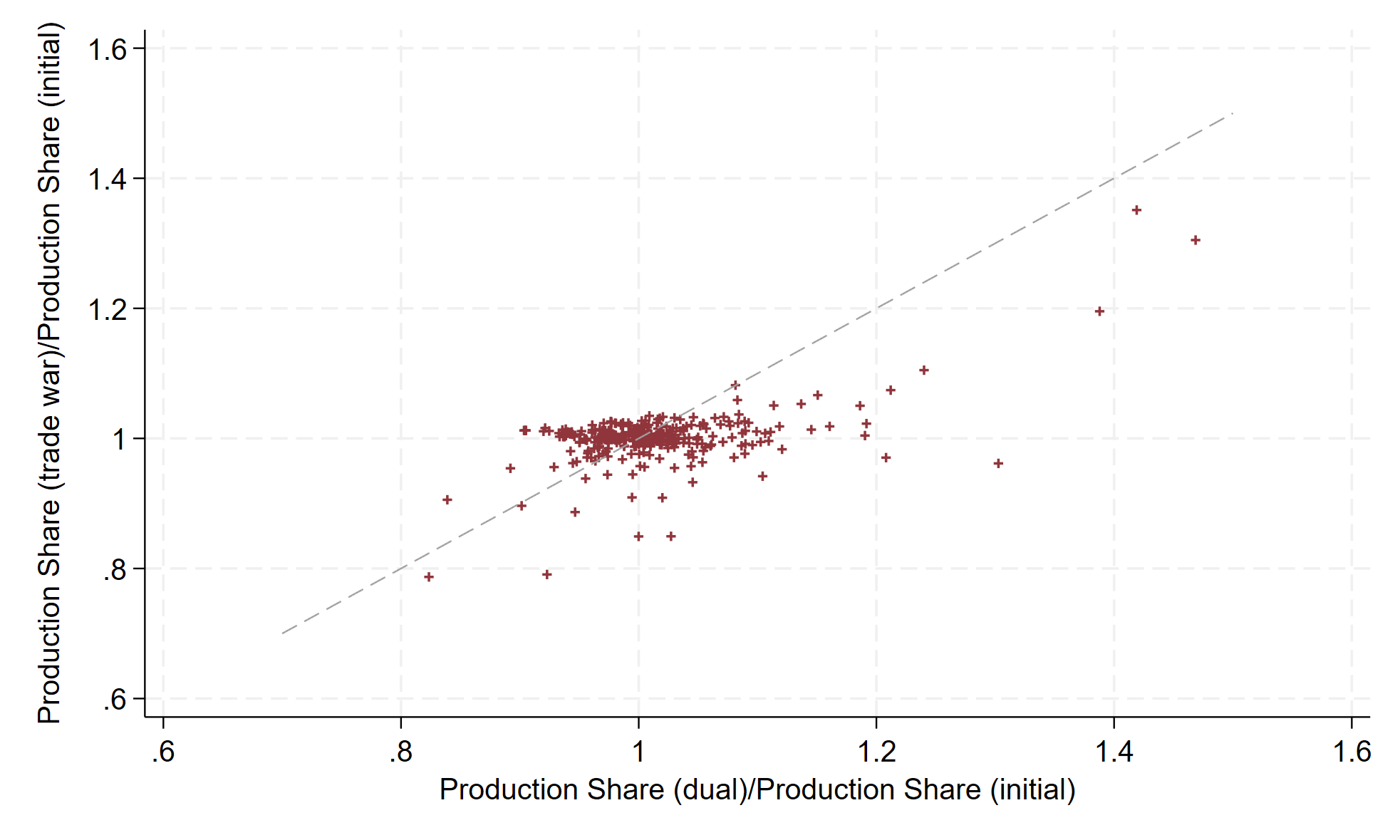}
		
		\caption{Effects of Nash Tariffs and Dual Policies on Sectoral Production Shares}
		\label{fig:production_ratio_nash}
		
		\vspace{0.2cm}
		\scriptsize
		(Notes: Sectoral production share is defined as $\frac{\sum_{n=1}^{N}X_{in}^j}{\sum_{s=1}^{J}\sum_{n=1}^{N}X_{in}^s}$. Non-tradable sectors are excluded. The dash line is the 45-degree line.)
	\end{figure}

	How do Nash tariffs and dual policies, summarized in Table \ref{tab:nash_theta_psi}, affect labor allocation across sectors? Figure \ref{fig:production_ratio_nash} compares changes in production shares under a global trade war and under dual competition. The results indicate that dual competition significantly impacts the production shares of most tradable sectors more than the trade war. This implies that, compared to dual policies, tariffs are a less effective tool for shifting labor towards sectors with large economies of scale.
	
\begin{figure}[htbp]
		\centering
		\includegraphics[width=0.5\textwidth]{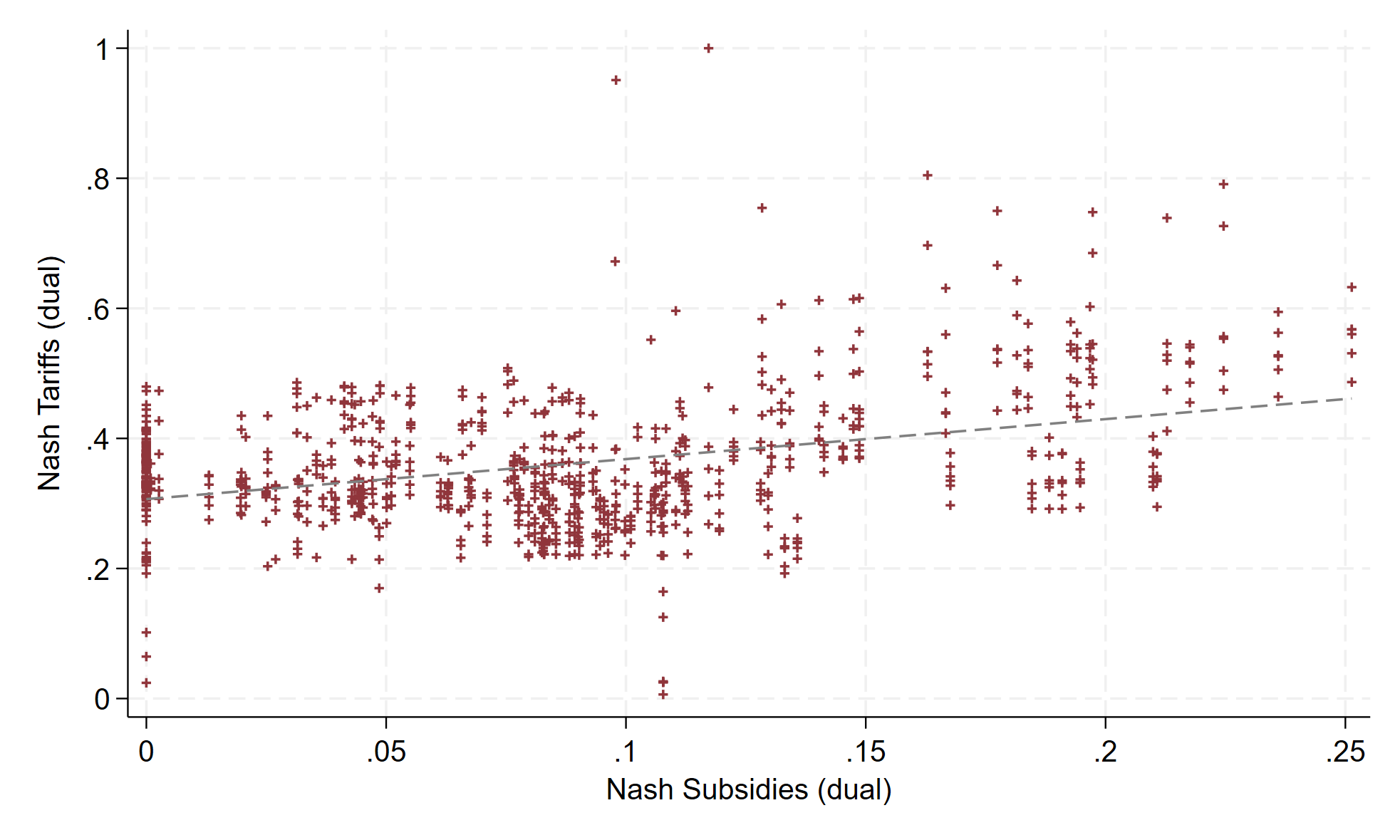}
		
		\caption{Nash Tariffs and Subsidies under Dual Competition}
		\label{fig:tar_sub_compare}
		
		\vspace{0.2cm}
		\scriptsize
		(Notes: Regressing Nash tariffs on Nash subsidies (controlling for importer fixed effects) leads to a coefficient $0.61$ with standard error $0.05$.)
	\end{figure}

	Table \ref{tab:nash_theta_psi} also indicates connections between Nash tariffs and industrial subsidies. To see it more clearly, we depict Nash tariffs and industrial subsidies in Figure \ref{fig:tar_sub_compare} and find a significantly positive correlation between these two policy tools. There are two driving forces behind this positive correlation. First, scale elasticities push Nash tariffs and industrial policies under dual competition towards the same direction (comparing column (1) with (2)  in Table \ref{tab:nash_theta_psi}). Second, $\theta_j$ and $\psi_j$ is negatively correlated under the calibration in \citet{Lashkaripour2019} (see Appendix Table \ref{tab: calibrate_lit2}). Smaller $\theta_j$ results in larger Nash tariffs and, thereby, the positive correlation between Nash tariffs and industrial subsidies.
	
	\subsection{Trade and Industrial Policy Competitions}
	\label{sec:welfare_nash}
	
\begin{figure}[htbp]
		\centering
		\includegraphics[width=0.66\textwidth]{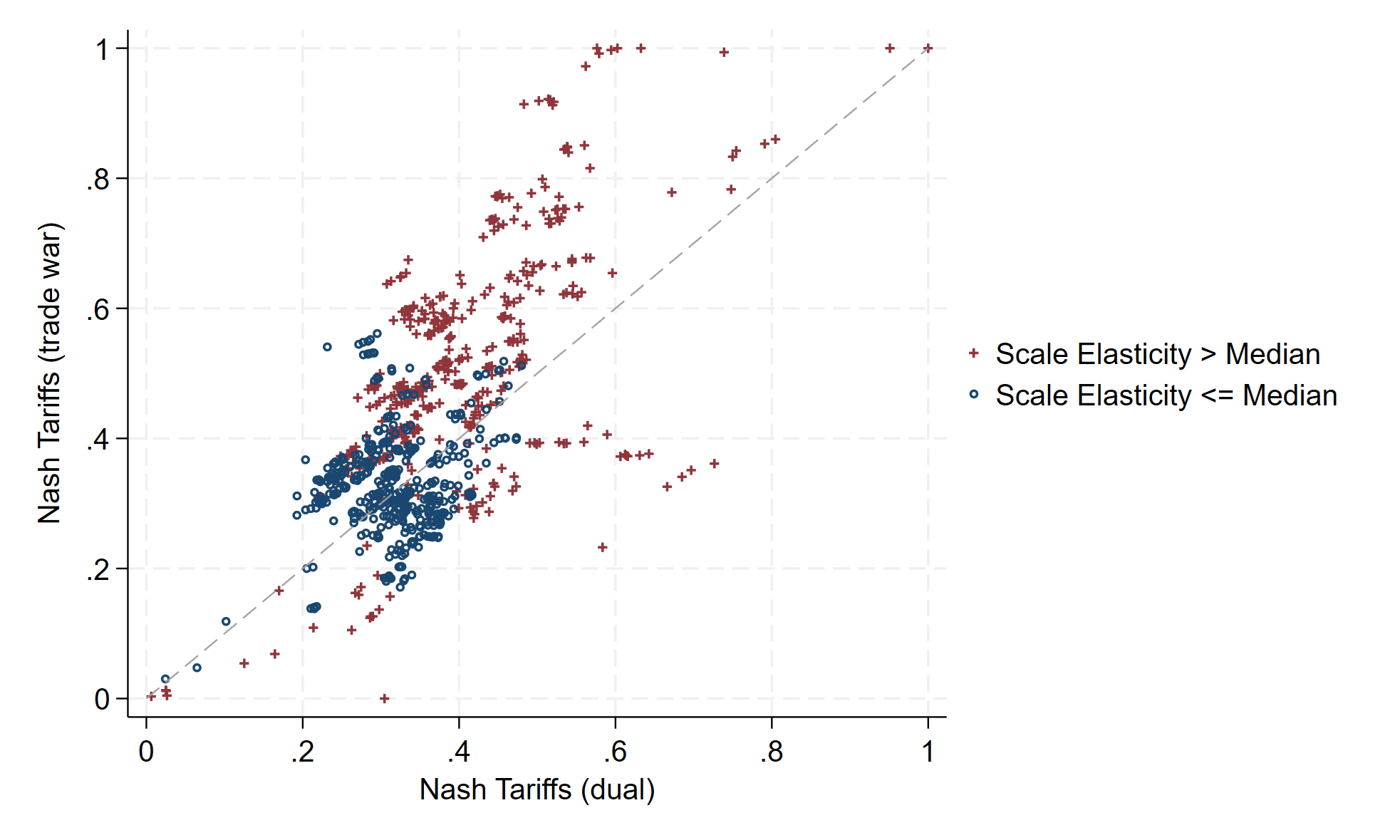}
		
		\caption{Nash Tariffs under Trade War vs. under Dual Competition}
		\label{fig:tar_nash_compare}
		
		\vspace{0.2cm}
		\scriptsize
		(Notes: Simple average of Nash tariffs under trade war is $42\%$, whereas simple average of Nash tariffs under dual competition is $35\%$.)
	\end{figure}

	In this section, we investigate the interactions of trade and industrial policies in global competition. To this end, we compare Nash tariffs under a trade war with those under dual competition. The results are depicted in Figure \ref{fig:tar_nash_compare}. It suggests that when industrial policy competition is allowed, countries have much weaker incentives to impose prohibitive non-cooperative tariffs: the simple average of Nash tariffs under dual competition is $35\%$, much lower than $42\%$ under a trade war. As shown in Table \ref{tab:nash_theta_psi}, Nash tariffs have a much larger positive correlation with scale elasticity $\psi_j$ under trade war than that under dual competition. Figure \ref{fig:tar_nash_compare} confirms this pattern: Countries impose prohibitive Nash tariffs on industries with $\psi_j$ above the median under a trade war. A policy implication of this result is that when industrial policies are not considered, the global competition tends to result in high tariffs on industries with strong increasing returns to scale. Recent examples in the US-China trade war starting from 2018 include semiconductors, electric vehicles, and photovoltaics.
	
	What are the welfare consequences of global competition when tariffs and industrial policies are combined in the policy consideration? Table \ref{table:nash_wefare_scale2} reports the welfare effects of the US-China Nash policies and of the global Nash policies. The results suggest that, compared with trade wars, dual competition via the combination of tariffs and industrial policies leads to larger welfare gains (or lower welfare losses). If a country is able to efficiently subsidize its high-tech industries with strong economies of scale, it has lower incentives to impose high import tariffs on these industries, as shown in Figure \ref{fig:tar_nash_compare}. As a result, dual competition tends to result in fewer distortions and, thereby, better welfare outcomes than tariff wars (Column (4) in Table \ref{table:nash_wefare_scale2}). 

	\begin{table}[htbp]\footnotesize
		\centering
		\caption{Welfare Effects of Nash Policies}
		\label{table:nash_wefare_scale2}
		
		\begin{tabular}{lcccc}
			\toprule 
			& \multicolumn{2}{c}{\textbf{China and US ($\Delta\%$)}} & \multicolumn{2}{c}{\textbf{World ($\Delta\%$)}} \\\cmidrule(lr){2-3}\cmidrule(lr){4-5}
			 & Trade war & Dual & Trade war & Dual  \\
			& (1)    &   (2)   & (3)   & (4)   \\
			United States   & -0.07 & 1.23     & -1.01  & 0.26  \\
			China   & -0.33 & 3.17    & -2.56  & 0.52  \\
			European Union   & 0.02 & -0.49    & -1.80  & -0.27  \\
			Japan   & 0.03 & -0.57    & -2.21  & -0.42  \\
			India   & 0.01 & -0.51   & -1.84  & 0.27  \\
			Brazil   & 0.00 & -0.05    & -2.08  & 0.23 \\
			Rest of the World   & 0.05 & -0.41    & -2.33  & -0.62 \\
			\bottomrule
			
			\multicolumn{5}{l}{
				\begin{minipage}{10cm}
					\vspace{0.5\baselineskip}
					\scriptsize \textit{Note: ``Trade war" refers to cases where players can modify their import tariffs solely. ``Dual" refers to cases where players have the flexibility to adjust both their industry subsidies and import tariffs. ``China and US" refers to cases where only China and the US are allowed to adjust their policies, whereas ``World" refers to cases where all economies can adjust their policies.}       		
			\end{minipage}}
		\end{tabular}				
	\end{table}
	
	Comparing Columns (3) and (4) in Table \ref{table:nash_wefare_scale2}, we observe that industrial policy competition generally leads to welfare gains for participants.\footnote{Appendix Table \ref{apptable:nash_wefare_scale2} reports welfare effects of industrial subsidy competitions. The results show that all major economies gain from global industrial subsidy competition.} The positive welfare impact of non-cooperative industrial policies, achieved by directing labor into high-economy-of-scale sectors, outweighs the negative welfare impact caused by terms-of-trade deterioration, \emph{i.e.} the ``immiserizing effect" of non-cooperative industrial policies highlighted by \citet{Lashkaripour2019}.

	\subsection{Extension: Cooperative Policies}
	
	Having characterized non-cooperative policy competitions across major economies, we turn to consider cooperative policies aiming to maximize global welfare. In particular, we consider a world social planner that chooses trade and industrial policies in all economies to maximize the weighted-average welfare change defined as follows
	\begin{equation}
		\label{eq:W_coop}
		\hat{W}\equiv\sum_{n=1}^{N}\bar{\omega}_n\left(\frac{\hat{Y}_{n}}{\hat{P}_{n}}\right),\quad \bar{\omega}_n\equiv \frac{Y_n}{\sum_{k=1}^{N}Y_k}.
	\end{equation}
	
	Notice that (i) the global cooperative policies may lead to welfare reduction in certain countries since the global social planner cares about weighted-average welfare changes in the global economy; and (ii) since cooperative policies have to be determined simultaneously in all countries and sectors, solving for these policies are much more computationally challenging than solving for the Nash policies. It takes about 24 hours to solve the global dual cooperative policies in which the global social planner chooses tariffs and subsidies in all countries to maximize the welfare expressed by Equation \eqref{eq:W_coop}.
	
	Our first finding is that global cooperation leads to very low tariffs: the average tariff under trade cooperation is $3.31\%$, and it will reduce to $2.23\%$ if both tariffs and industrial policies are considered in global cooperation. Cooperative tariffs on many industries are close to zero. In sum, global cooperation tends to make the world tariff-free.
		
	We then investigate the correlation between cooperative policies and trade and scale elasticities. The regression results are reported in Table \ref{tab:coop_theta_psi}. We find that if tariff is the only policy tool for global cooperation, tariffs are lower in industries with higher scale elasticities ($\psi_j$) and lower trade elasticities ($\theta_j$). Cooperative tariffs thus reallocate factors of production from weak-increasing-return-to-scale industries to strong-increasing-return-to-scale industries and result in welfare gains. When industrial policies are used for global cooperation, this reallocation is mainly achieved by industrial subsidies, which are much larger in industries with higher $\psi_j$ and lower $\theta_j$.
	
	\begin{table}[htbp]\footnotesize
		\centering
		\caption{Cooperative Policies and Trade and Scale Elasticities}
		\label{tab:coop_theta_psi}
		\begin{tabular}{lccc}\toprule
			Dep. Var: & $\log\left(t_{in}^j\right)$-coop & $\log\left(t_{in}^j\right)$-dual coop  & $\log\left(\frac{1}{1-s_i^j}\right)$  \\\cmidrule(lr){2-4}
			& (1) & (2) & (3) \\
			$\log\left(\theta_j\right)$ & $1.83^{***}$ & $-0.33^{***}$ &   $-0.07^{***}$  \\
			& $(0.50)$ & $(0.10)$ &  $(0.01)$   \\
			$\log\left(\psi_j\right)$ &  $0.99$     &   $0.28^{**}$    & $0.13^{***}$ \\
			& $(0.60)$      &  $(0.13)$     & $(0.01)$  \\
			Exp. FE &  $\checkmark$     &   $\checkmark$    &  $\checkmark$     \\
			Imp. FE &  $\checkmark$     &   $\checkmark$    &      \\
			\# obs & 229   & 525   & 154    \\
			$R^2$  & 0.24  & 0.21  & 0.80  \\
			\bottomrule
		\end{tabular}
	\end{table}
	
	Finally, We quantify the welfare consequences of cooperative policies and report the results in Table \ref{table:nash_wefare_scale_coop}. Cooperative policies improve welfare in most of the major economies, except for India under dual cooperation. For all major economies except India, dual cooperation results in larger welfare gains than tariff cooperation. Comparing the welfare effects reported in Table \ref{table:nash_wefare_scale2} for Nash policies and in Table \ref{table:nash_wefare_scale_coop} for cooperative policies, we can figure out large welfare gains by pushing the world from non-cooperative competition to global cooperation.	
	
	\begin{table}[htbp]\footnotesize
		\centering
		\caption{Welfare Effects of Cooperative Policies ($\Delta\%$)}
		\label{table:nash_wefare_scale_coop}
		
		\begin{tabular}{lcc}
			\toprule
			& Tariff & Dual\\\cmidrule(lr){2-3}
			& (1)    & (2)  \\
			United States   & 0.19  & 2.82 \\
			China   & 1.13  & 4.29 \\
			European Union   & 0.19  & 1.78 \\
			Japan   & 0.37  & 1.39 \\
			India   & 2.34  & -0.27 \\
			Brazil   & 0.08  & 3.46\\
			Rest of the World   & 0.06  & 1.88 \\
			\bottomrule
		\end{tabular}
	\end{table}
	
	\subsection{Extension: Imperfect Implementation of Industrial Subsidies}
	
	In section \ref{sec:welfare_nash}, we have shown that most of the major economies, including China, gain considerably from non-cooperative industrial subsidies. However, this welfare gain hinges on accurately specifying and implementing industrial subsidies. In this subsection, we consider the case in which China cannot perfectly implement its optimal industrial subsidies in the global industrial policy competition. Instead, China chooses from $\left[0.1 s_{CHN}^{j^*},1.9 s_{CHN}^{j^*}\right]$, following a uniform distribution, for all tradable sector $j$. 
	
\begin{figure}[htbp]
		\centering
		\includegraphics[width=0.5\textwidth]{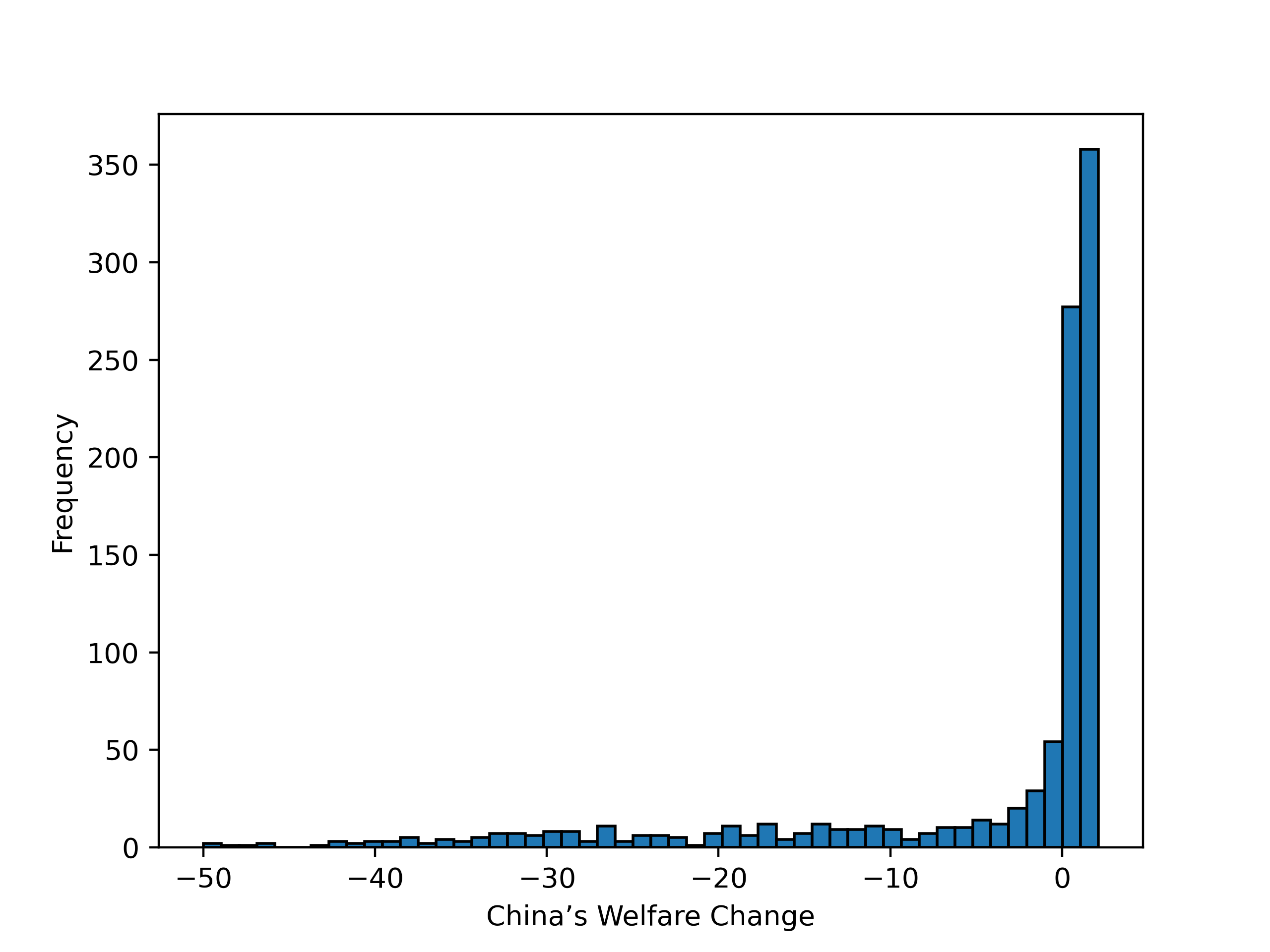}
		
		\caption{China's Welfare Change (\%) under Randomly Drawn Industrial Subsidies}
		\label{fig:welfare_cn_random}
		
		\vspace{0.2cm}
		\footnotesize
		(Note: Each $s_{CHN}^j$ is drawn uniformly from $\left[0.1 s_{CHN}^{j^*},1.9 s_{CHN}^{j^*}\right]$. We draw $\left(s_{CHN}^j\right)_{j=1}^{22}$ for 1000 times.)
	\end{figure}

	We randomly draw China's subsidies 1000 times. Figure \ref{fig:welfare_cn_random} is the histogram of the corresponding welfare changes in China. It suggests that given other countries set their industrial subsidies optimally; China has to precisely specify its industrial subsidies to gain from global industrial policy competition.\footnote{See Appendix Table \ref{apptable:nash_wefare_scale2} for welfare effects of global industrial policy competition.} Once deviate from its optimal values, China's industrial subsidies tend to result in small welfare gains or, in many cases, welfare losses.
	
	\section{Conclusion\label{sec:Conclusion}}
	
	In this paper, we introduce a deep learning framework, DL-opt, designed to determine optimal policies within quantitative trade and spatial models. This framework integrates the nested fixed point (NFXP) algorithm, automatic implicit differentiation, and best-response dynamics. Our approach efficiently computes high-dimensional optimal policies in trade models with intricate equilibrium systems, making it particularly suitable for analyzing the incentives and consequences of contemporary global trade and industrial policy competitions.
	
	We apply our framework to calculate optimal trade and industrial policies within a multi-country-multi-sector general equilibrium trade model that incorporates input-output linkages and sectoral scale economies. Our counterfactual analysis highlights the importance of sectoral heterogeneity and policy combinations in understanding global economic competition.
	
	Our method is broadly applicable to trade and spatial economics. It effectively overcomes the curse of dimensionality, enabling the computation of high-dimensional optimal (continuous) policies in models where numerous agents, potentially regions, interact. Applications of this method include but are not limited to, optimizing carbon emissions, corporate taxes, and innovation subsidies across different spatial contexts.
	
	\bibliographystyle{ecta}
	\bibliography{reference}

\begin{thebibliography}{17}
\newcommand{\enquote}[1]{``#1''}
\expandafter\ifx\csname natexlab\endcsname\relax\def\natexlab#1{#1}\fi

\bibitem[\protect\citeauthoryear{Abadi, Agarwal, Barham, Brevdo, Chen, Citro, Corrado, Davis, Dean, Devin et~al.}{Abadi et~al.}{2016}]{AAB+2016}
\textsc{Abadi, M., A.~Agarwal, P.~Barham, E.~Brevdo, Z.~Chen, C.~Citro, G.~S. Corrado, A.~Davis, J.~Dean, M.~Devin, et~al.} (2016): \enquote{TensorFlow: Large-scale machine learning on heterogeneous distributed systems,} \emph{arXiv preprint arXiv:1603.04467}.

\bibitem[\protect\citeauthoryear{Bartelme, Costinot, Donaldson, and Rodriguez-Clare}{Bartelme et~al.}{2021}]{Bartelme2019}
\textsc{Bartelme, D., A.~Costinot, D.~Donaldson, and A.~Rodriguez-Clare} (2021): \enquote{The Textbook Case for Industrial Policy: Theory Meets Data,} \emph{Mimeo}.

\bibitem[\protect\citeauthoryear{Dekle, Eaton, and Kortum}{Dekle et~al.}{2008}]{DEK2008}
\textsc{Dekle, R., J.~Eaton, and S.~Kortum} (2008): \enquote{Global rebalancing with gravity: Measuring the burden of adjustment,} \emph{IMF Staff Papers}, 55, 511--540.

\bibitem[\protect\citeauthoryear{Eaton and Kortum}{Eaton and Kortum}{2002}]{Eaton2002}
\textsc{Eaton, J. and S.~Kortum} (2002): \enquote{Technology, Geography, and Trade,} \emph{Econometrica}, 70, 1741--1779.

\bibitem[\protect\citeauthoryear{Feng, Han, and Zhu}{Feng et~al.}{2023}]{Feng2023}
\textsc{Feng, Z., J.~Han, and S.~Zhu} (2023): \enquote{Optimal Taxation with Incomplete Markets: an Exploration via Reinforcement Learning,} \emph{Mimeo}.

\bibitem[\protect\citeauthoryear{Frostig, Johnson, and Leary}{Frostig et~al.}{2018}]{FJL2018}
\textsc{Frostig, R., M.~J. Johnson, and C.~Leary} (2018): \enquote{Compiling machine learning programs via high-level tracing,} \emph{Systems for Machine Learning}, 4.

\bibitem[\protect\citeauthoryear{Han, E, and Yang}{Han et~al.}{2021}]{Han2021}
\textsc{Han, J., W.~E, and Y.~Yang} (2021): \enquote{DeepHAM: A Global Solution Method for Heterogeneous Agent Models with Aggregate Shocks,} \emph{Mimeo}.

\bibitem[\protect\citeauthoryear{Heinrich, Jang, Mungo, Pangallo, Scott, Tarbush, and Wiese}{Heinrich et~al.}{2023}]{HJMP+2021}
\textsc{Heinrich, T., Y.~Jang, L.~Mungo, M.~Pangallo, A.~Scott, B.~Tarbush, and S.~Wiese} (2023): \enquote{Best-response dynamics, playing sequences, and convergence to equilibrium in random games,} \emph{International Journal of Game Theory}, 1--33.

\bibitem[\protect\citeauthoryear{Ju, Ma, Wang, and Zhu}{Ju et~al.}{2024}]{JMWZ2023}
\textsc{Ju, J., H.~Ma, Z.~Wang, and X.~Zhu} (2024): \enquote{Trade Wars and Industrial Policy Competitions: Understanding the US-China economic conflicts,} \emph{Journal of Monetary Economics}, 141.

\bibitem[\protect\citeauthoryear{Judd and Su}{Judd and Su}{2012}]{Judd2012}
\textsc{Judd, K.~L. and C.-L. Su} (2012): \enquote{Constrained Optimization Approaches to Estimation of Structural Models,} \emph{Econometrica}, 80.

\bibitem[\protect\citeauthoryear{Kingma and Ba}{Kingma and Ba}{2014}]{KB2014}
\textsc{Kingma, D.~P. and J.~Ba} (2014): \enquote{Adam: A method for stochastic optimization,} \emph{arXiv preprint arXiv:1412.6980}.

\bibitem[\protect\citeauthoryear{Lashkaripour and Lugovskyy}{Lashkaripour and Lugovskyy}{2023}]{Lashkaripour2019}
\textsc{Lashkaripour, A. and V.~Lugovskyy} (2023): \enquote{Profits, scale economies, and the gains from trade and industrial policy,} \emph{American Economic Review}, 113, 2759--2808.

\bibitem[\protect\citeauthoryear{Ossa}{Ossa}{2014}]{Ossa2014}
\textsc{Ossa, R.} (2014): \enquote{Trade Wars and Trade Talks with Data,} \emph{American Economic Review}, 104, 4104--4146.

\bibitem[\protect\citeauthoryear{Paszke, Gross, Massa, Lerer, Bradbury, Chanan, Killeen, Lin, Gimelshein, Antiga et~al.}{Paszke et~al.}{2019}]{PGM+2019}
\textsc{Paszke, A., S.~Gross, F.~Massa, A.~Lerer, J.~Bradbury, G.~Chanan, T.~Killeen, Z.~Lin, N.~Gimelshein, L.~Antiga, et~al.} (2019): \enquote{PyTorch: An imperative style, high-performance deep learning library,} \emph{Advances in neural information processing systems}, 32.

\bibitem[\protect\citeauthoryear{Rust}{Rust}{2000}]{Rust2000}
\textsc{Rust, J.} (2000): \enquote{Nested Fixed Point Algorithm,} \emph{mimeo}.

\bibitem[\protect\citeauthoryear{Sun}{Sun}{2023}]{Sun2023}
\textsc{Sun, J.} (2023): \enquote{Continuation Value Is All You Need: A Deep Learning Method for Solving Heterogeneous-Agent Models with Aggregate Uncertainty,} \emph{Mimeo}.

\bibitem[\protect\citeauthoryear{Zhao, Zhou, Li, Tang, Wang, Hou, Min, Zhang, Zhang, Dong et~al.}{Zhao et~al.}{2023}]{ZZL+2023}
\textsc{Zhao, W.~X., K.~Zhou, J.~Li, T.~Tang, X.~Wang, Y.~Hou, Y.~Min, B.~Zhang, J.~Zhang, Z.~Dong, et~al.} (2023): \enquote{A survey of large language models,} \emph{arXiv preprint arXiv:2303.18223}.

\end{thebibliography}
	
	\newpage 
	\appendix
	
	\setcounter{figure}{0}	\renewcommand{\thefigure}{\Alph{section}.\arabic{figure}}
	\setcounter{table}{0}
	\renewcommand{\thetable}{\Alph{section}.\arabic{table}}
	\renewcommand{\theequation}{\Alph{section}.\arabic{equation}}
	\setcounter{equation}{0}
	
	\section{Model, Calibration, and Computation}
	
	\subsection{Automatic Differentiation: A Toy Example}
	\label{appsec:AD}
	
	We consider function $y = f\left(x_1,x_2\right) = \left[x_1^2 + x_1/x_2 - \exp\left(x_2\right)\right]\left[x_1/x_2-\exp\left(x_2\right)\right]$, aiming to compute $\frac{\partial y}{\partial x_i}$ at point $\left(\bar{x}_1,\bar{x}_2\right)$. 
	
	We can use finite difference to approximate $\frac{\partial y}{\partial x_1}$: $\frac{\partial y}{\partial x_1}\simeq \frac{f(\bar{x}_1+h,\bar{x}_2)-f(\bar{x}_1,\bar{x}_2)}{h}$. Notice that to compute $\frac{\partial y}{\partial x_1}$ and $\frac{\partial y}{\partial x_2}$, we need to evaluate function $f\left(x_1,x_2\right)$ at least three times, including initial evaluation of $f\left(\bar{x}_1,\bar{x}_2\right)$.
	
	Automatic differentiation, instead, introduces a series of intermediate variables: $v_{-1}=x_1$, $v_{0}=x_2$, $v_1=v_{-1}/v_0$, $v_2 = v_{-1}^2$, $v_3=\exp\left(v_0\right)$, $v_4=v_1-v_3$, $v_5=v_2+v_4$, and $v_6=v_4\cdot v_5 = y$. The relationship between $\left(x_1,x_2\right)$ and $y$, through these intermediate variables, can be illustrated by the computation graph in Figure \ref{fig:cg}:
	\begin{figure}[htbp]
		\centering
		\includegraphics[width=0.4\textwidth]{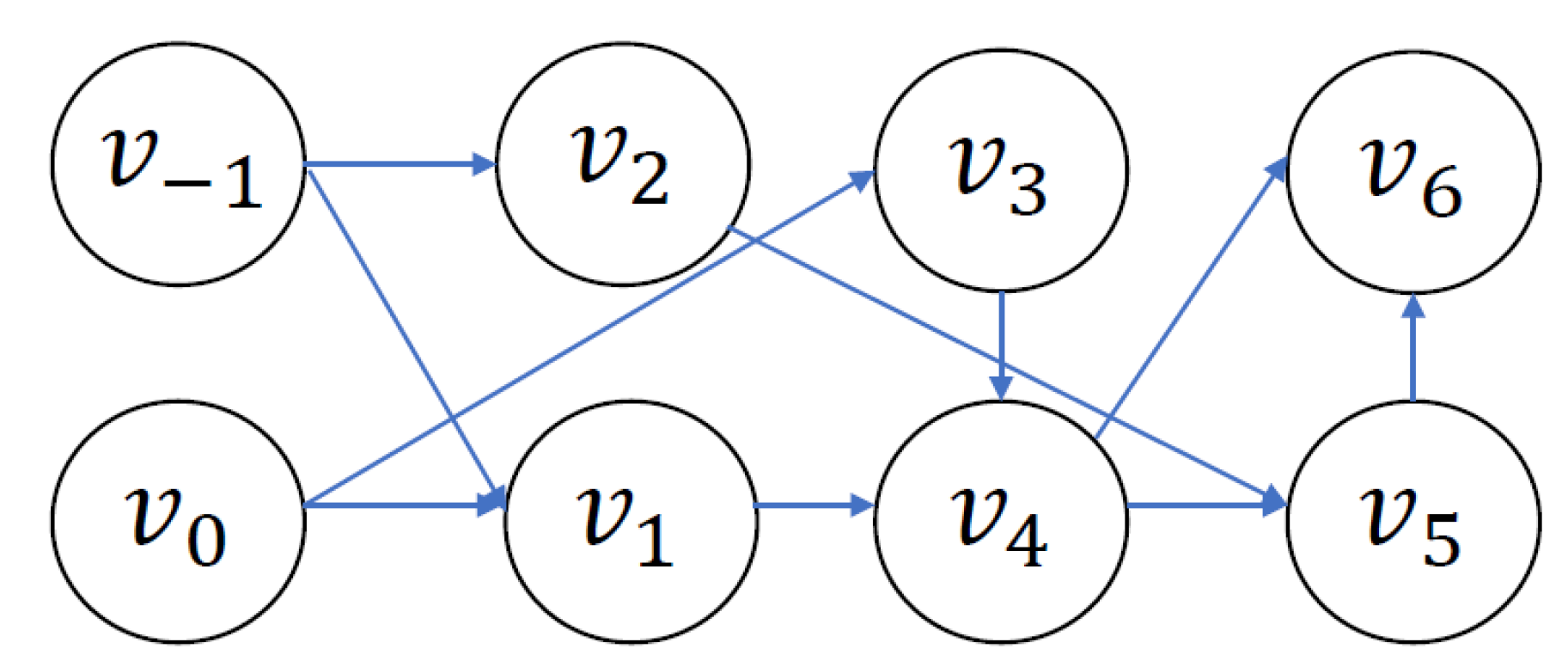}
		\caption{Computation Graph of Automatic Differentiation: Toy Example}
		\label{fig:cg}
	\end{figure}
	
	By the chain rule, we have
	\begin{equation}
		\begin{aligned}
			&\frac{\partial y}{\partial x_1} = \frac{\partial y}{\partial v_6}\left(\frac{\partial v_6}{\partial v_4}\frac{\partial v_4}{\partial v_1}\frac{\partial v_1}{\partial v_{-1}}+\frac{\partial v_6}{\partial v_5}\left(\frac{\partial v_5}{\partial v_4}\frac{\partial v_4}{\partial v_1}\frac{\partial v_1}{\partial v_{-1}}+\frac{\partial v_5}{\partial v_2}\frac{\partial v_2}{\partial v_{-1}}\right)\right)\frac{\partial v_{-1}}{\partial x_1},\\
			&\frac{\partial y}{\partial x_2} = \frac{\partial y}{\partial v_6}\left(\frac{\partial v_6}{\partial v_4}\left(\frac{\partial v_4}{\partial v_1}\frac{\partial v_1}{\partial v_0}+\frac{\partial v_4}{\partial v_3}\frac{\partial v_3}{\partial v_0}\right)+\frac{\partial v_6}{\partial v_5}\frac{\partial v_5}{\partial v_4}\frac{\partial v_4}{\partial v_1}\frac{\partial v_1}{\partial v_0}\right)\frac{\partial v_{0}}{\partial x_2}.
		\end{aligned}
	\end{equation}
	
	Notice that there are many common terms in computing $\frac{\partial y}{\partial x_1}$ and $\frac{\partial y}{\partial x_2}$. Therefore, we do not need to evaluate $f\left(x_1,x_2\right)$ twice. We only need to compute the following terms: $\frac{\partial y}{\partial v_6}$, $\frac{\partial v_{-1}}{\partial x_1}$, $\frac{\partial v_0}{\partial x_2}$, $\frac{\partial v_6}{\partial v_4}$, $\frac{\partial v_6}{\partial v_5}$, $\frac{\partial v_5}{\partial v_4}$, $\frac{\partial v_5}{\partial v_2}$, $\frac{\partial v_4}{\partial v_3}$, $\frac{\partial v_4}{\partial v_1}$, $\frac{\partial v_3}{\partial v_0}$, $\frac{\partial v_2}{\partial v_{-1}}$, $\frac{\partial v_1}{\partial v_0}$, $\frac{\partial v_1}{\partial v_{-1}}$. Calculating the gradients for these elementary functions is much more efficient than evaluating the entire function multiple times.
	
	\subsection{``Exact-hat" algebra}
	\label{sec:exact-hat}
	
	Changes in unit costs can be expressed as
	\begin{equation}
		\label{eq: hatc}
		\hat{c}^j_i = \frac{1}{\left(\hat{L}^j_i\right)^{\psi_j}}\hat{w_i}^{\beta^j_i}\left[\prod_{s=1}^{J}\left(\hat{P}^s_i\right)^{\gamma^{sj}_i}\right]^{1-\beta^j_i}.
	\end{equation}
	
	Changes in trade share:
	\begin{equation}
		\label{eq:hatpi}
		\hat{\pi}^j_{in} = \frac{\left[\hat{c}^j_i\widehat{1+t^j_{in}}\widehat{1+e^j_{in}}\right]^{-\theta_j}}{\left(\hat{P}^j_n\right)^{-\theta_j}}.
	\end{equation}
	
	Changes in price indices:
	\begin{equation}
		\label{eq:hatP}
		\hat{P}^j_n = \left[\sum_{i=1}^N \pi^j_{in}\left[\hat{c}^j_i\widehat{1+t^j_{in}}\widehat{1+e^j_{in}}\right]^{-\theta_j}\right]^{-\frac{1}{\theta_j}}.
	\end{equation}
	
	Changes in sectoral wage incomes:
	\begin{equation}
		\label{eq:hatw}
		\hat{w}_i\hat{L}_i^jw_iL^j_i = \beta^j_i\sum_{n=1}^{N}\frac{\hat{\pi}^j_{in}\hat{X}^j_nX^j_{in}}{\left(1+t^j_{in}\right)'\left(1+e^j_{in}\right)'}.
	\end{equation}
	
	Changes in sectoral labor allocation satisfy:
	\begin{equation}
		\label{eq:hatL}
		\sum_{j=1}^{J}\hat{L}_i^jL_i^j = \bar{L}_i.
	\end{equation}
	
	Changes in the total income:
	\begin{equation}
		\label{eq:hatY}
		\begin{aligned}
			\hat{Y}_iY_i = \hat{w}_iw_i\bar{L}_i+\sum_{j=1}^J\sum_{n=1}^{N}\frac{\left(e^j_{in}\right)'}{\left(1+e^j_{in}\right)'}\left(X^j_{in}\right)'+\sum_{j=1}^{J}\sum_{k=1}^{N}\frac{\left(t^j_{ki}\right)'}{\left(1+t^j_{ki}\right)'\left(1+e^j_{ki}\right)'}\left(X^j_{ki}\right)'.
		\end{aligned}
	\end{equation}
	
	Changes in sectoral expenditure:
	\begin{equation}
		\label{eq: hatX}
		\begin{aligned}
			\hat{X}^j_iX^j_i = \alpha^j_i\hat{Y}_iY_i + \sum_{s=1}^{J}\left(1-\beta^s_i\right)\gamma^{js}_i\sum_{n=1}^{N}\frac{\left(X^s_{in}\right)'}{\left(1+t^s_{in}\right)'\left(1+e^s_{in}\right)'}.
		\end{aligned}
	\end{equation}

	Changes in aggregate price indices:
	\begin{equation}
		\hat{P}_n = \prod_{j=1}^{J}\left(\hat{P}^j_n\right)^{\alpha^j_n}.
	\end{equation}
	
	Optimal policies are solved by
	\begin{equation}
		\begin{aligned}
			&\max_{\left\{t_{in}^{j'},e_{ni}^{j'}; \hat{w}_{i},\hat{L}_{i}^{j},\hat{P}_{i}^{j},\hat{X}_{i}^{j}\right\}_{i,j}}\hat{W}_n\equiv\frac{\hat{Y}_{n}}{\hat{P}_{n}},\quad\forall n=1,2,\cdots,N\\
			&\text{s.t. Equation \eqref{eq:hatP}, \eqref{eq:hatL},\eqref{eq:hatw}, and \eqref{eq: hatX}}
		\end{aligned}
	\end{equation}

	\subsection{Calibration of $\left(\psi_j,\theta_j\right)$}
	\label{appsec:calibration}
	
	The sector-specific trade and scale elasticities, $\left(\theta_j,\psi_j\right)$, are calibrated from \citet{Lashkaripour2019}. The calibrated values are summarized in Table \ref{tab: calibrate_lit2}.
	
        \begin{table}[htbp]\scriptsize
		\centering
		\caption{Calibration of $\left(\psi_j, \theta_j\right)$ from \citet{Lashkaripour2019}}
		\label{tab: calibrate_lit2}
		
		\begin{tabular}{lllcc}\toprule
			Industry & ICIO code & Description & $\theta_j$ & $\psi_j$   \\\midrule
			1  & D01T02 & Agriculture               & 6.23  & 0.14 \\
			2  & D03    & Fishing                   & 6.23  & 0.14 \\
			3  & D05T06 & Mining, energy            & 5.28  & 0.17 \\
			4  & D07T08 & Mining, non-energy        & 5.28  & 0.17 \\
			5  & D09    & Mining support            & 5.28  & 0.17 \\
			6  & D10T12 & Food                      & 2.30  & 0.35 \\
			7  & D13T15 & Textiles                  & 3.36  & 0.22 \\
			8  & D16    & Wood                      & 3.90  & 0.23 \\
			9  & D17T18 & Paper                     & 2.65  & 0.32 \\
			10 & D19    & Petroleum                 & 0.64  & 0.35 \\
			11 & D20    & Chemical                  & 3.97  & 0.23 \\
			12 & D21    & Pharmaceutical            & 3.97  & 0.23 \\
			13 & D22    & Rubber                    & 5.16  & 0.14 \\
			14 & D23    & Non-metallic              & 5.28  & 0.17 \\
			15 & D24    & Basic metals              & 3.00  & 0.21 \\
			16 & D25    & Fabricated metal          & 3.00  & 0.21 \\
			17 & D26    & Computer                  & 1.24  & 0.55 \\
			18 & D27    & Electrical equipment      & 1.24  & 0.55 \\
			19 & D28    & Machinery nec             & 7.75  & 0.12 \\
			20 & D29    & Motor vehicles            & 2.81  & 0.13 \\
			21 & D30    & Other transport equipment & 2.81  & 0.13 \\
			22 & D31T33 & Manufacturing nec         & 6.17  & 0.15 \\
			\bottomrule
			\multicolumn{5}{l}{
				\begin{minipage}{14cm}
					\vspace{0.5\baselineskip}
					\scriptsize \textit{Notes:}  We set $\theta_j = 10$ and $\psi_j = 0$ for non-tradable sectors.      		
			\end{minipage}}
		\end{tabular}
	\end{table}
	
	\subsection{Adaptive Moment Estimation (ADAM) Algorithm}
	\label{appsec:adam}
	
	The Adaptive Moment Estimation (ADAM, see \cite{KB2014}) algorithm is an extension of gradient descent that combines adaptive learning rates and momentum. It utilizes the first and second moments of the gradients to adaptively adjust the learning rate for each parameter. ADAM is widely employed in machine learning optimization, such as large language models (LLMs) GPT3 (175B) with 175 billions parameters, BLOOM (176B), MT-NLG (530B), Gopher (280B), ERNIE 3.0 Titan (260B), and so on (see e.g. the survey \cite{ZZL+2023}).
	
	The update step in ADAM involves the following steps:
	\begin{enumerate}
		\item Compute the gradient of the objective function with respect to
		the parameters: $\nabla L_{n}|_{a=a^{t}}$. 
		\item Calculate the first moment estimate of the gradients, $m_{t}=\beta_{1}m_{t-1}+(1-\beta_{1})\nabla L_{n}|_{a=a^{t}}$, where $\beta_{1}$ is the first moment decay rate and $m_{0}=0$.
		\item Calculate the second moment estimate of the gradients $v_{t}=\beta_{2}v_{t-1}+(1-\beta_{2})(\nabla L_{n}|_{a=a^{t}})^{\odot2}$, where $\beta_{2}$ is the second moment decay rate, $v_{0}=0$, and $(\nabla L_{n}|_{a=a^{t}})^{\odot2}$ is the element-wise square.
		\item Bias-correct the first and second moment estimates, $\hat{m}_{t}=\frac{m_{t}}{1-\beta_{1}^{t}},\ \ \hat{v}_{t}=\frac{v_{t}}{1-\beta_{2}^{t}}$, where $\beta_{1}^{t}$ and $\beta_{2}^{t}$ are values $\beta_{1}$ and $\beta_{2}$ to the power $t$.
		\item Update the parameters using the bias-corrected estimates: 		
		\begin{equation}
			a^{t+1}=a^{t}-\gamma\frac{\hat{m}_{t}}{\sqrt{\hat{v}_{t}}+\epsilon},
		\end{equation}
		where $\epsilon$ is a small constant to avoid division by zero.
	\end{enumerate}
	
	In the above steps, $a^{t}$ represents the parameter value at iteration	$t$, $\gamma$ is the learning rate, $\beta_{1}$ and $\beta_{2}$	are the decay rates for the first and second moments, and $m_{t}$ and $v_{t}$ are the first and second moment estimates at iteration $t$, respectively. For our best-response search in each round, we utilize the ADAM optimization method, which offers faster convergence compared to naive gradient descent.

	\setcounter{figure}{0}	\renewcommand{\thefigure}{\Alph{section}.\arabic{figure}}
	\setcounter{table}{0}
	\renewcommand{\thetable}{\Alph{section}.\arabic{table}}
	\renewcommand{\theequation}{\Alph{section}.\arabic{equation}}
	\setcounter{equation}{0}
	
	\newpage
	\section{Quantification}
	
	\subsection{The ``Internal Cooperation" Assumption in \citet{Lashkaripour2019}}
	\label{appsec:internal_cooperation}
		
\begin{figure}[htbp]
		\centering
		\includegraphics[width=0.5\textwidth]{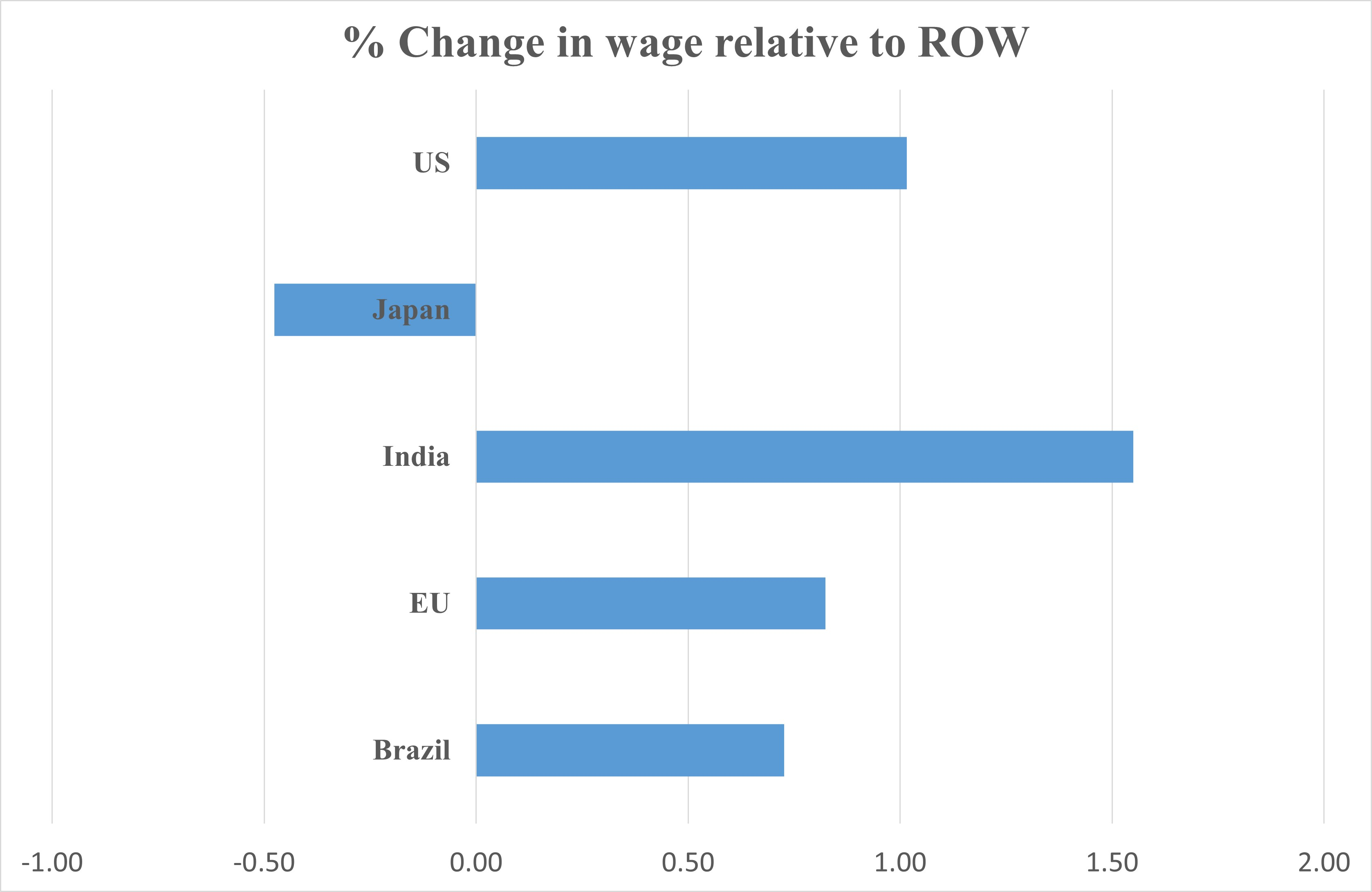}
		
		\caption{Changes in Relative Wage under China's Unilaterally Optimal Policies (in percentage)}
		\label{fig:China_full_hw}
		
		\vspace{0.2cm}
		\scriptsize
		(Notes: Here, to compare with the results in \citet{Lashkaripour2019}, we consider China's unilaterally optimal import tariffs, export tariffs, and industrial subsidies. We normalize the change in nominal wage in the rest of the world (ROW) as 0.)
	\end{figure}

	Figure \ref{fig:China_full_hw} suggests that the ``internal cooperation" assumption made by \citet{Lashkaripour2019} is restrictive in solving for the fully optimal policies: the relative wages among other major economies do change under China's unilaterally optimal policies. The changes in relative wage are moderate here because the economies considered in our numerical exercises are very large, the 6 major economies plus the rest of the world. In this case, the extraterritorial terms-of-trade effects are small. In a world with more, smaller, and more heterogeneous economies, the extraterritorial terms-of-trade effects of optimal policies would be larger, resulting in larger gaps between the theoretical optimal policies in \citet{Lashkaripour2019} and the fully optimal policies in our numerical exercises.
	
	\subsection{Welfare Effects of Industrial Subsidy Competition}
	
	This section reports welfare effects of industrial subsidy competitions, as a supplement to the welfare results in Table \ref{table:nash_wefare_scale2}.
	
	\begin{table}[htbp]\footnotesize
		\centering
		\caption{Welfare Effects of Industrial Subsidy Competitions}
		\label{apptable:nash_wefare_scale2}
		
		\begin{tabular}{lccc}
			\toprule 
			& \textbf{China and US ($\Delta\%$)} & \multicolumn{2}{c}{\textbf{World ($\Delta\%$)}} \\\cmidrule(lr){2-2}\cmidrule(lr){3-4}
			& Subsidy & Subsidy & Subsidy-Uni   \\
			& (1)    &   (2)   & (3)    \\
			United States & 1.28    & 1.31  & 1.31   \\
			China & 3.31    & 2.42  & 1.80   \\
			European Union & -0.51    & 1.03  & 0.53  \\
			Japan & -0.61    & 1.14  & 0.30    \\
			India & -0.50   & 2.56  & 0.60    \\
			Brazil & -0.05    & 1.97  & 1.71  \\
			Rest of the World & -0.44    & 1.47  & 1.76   \\
			\bottomrule
			
			\multicolumn{4}{l}{
				\begin{minipage}{10cm}
					\vspace{0.5\baselineskip}
					\scriptsize \textit{Note: ``Subsidy" refers cases where players can adjust their industry subsidies only. ``Subsidy-uni" refers to the cases where each player can only choose a uniform subsidy rate for all manufacturing sectors (sector 6-22 in Table \ref{tab: calibrate_lit2}). ``China and US" refers to cases where only China and the US are allowed to adjust their policies, whereas ``World" refers to cases where all economies can adjust their policies.}       		
			\end{minipage}}
		\end{tabular}
		
	\end{table}
    
	\subsection{Optimal Trade and Industrial Policies without Scale Economies}
	\label{sec:noscale}
	
\begin{figure}
		\centering
		\includegraphics[width=\textwidth]{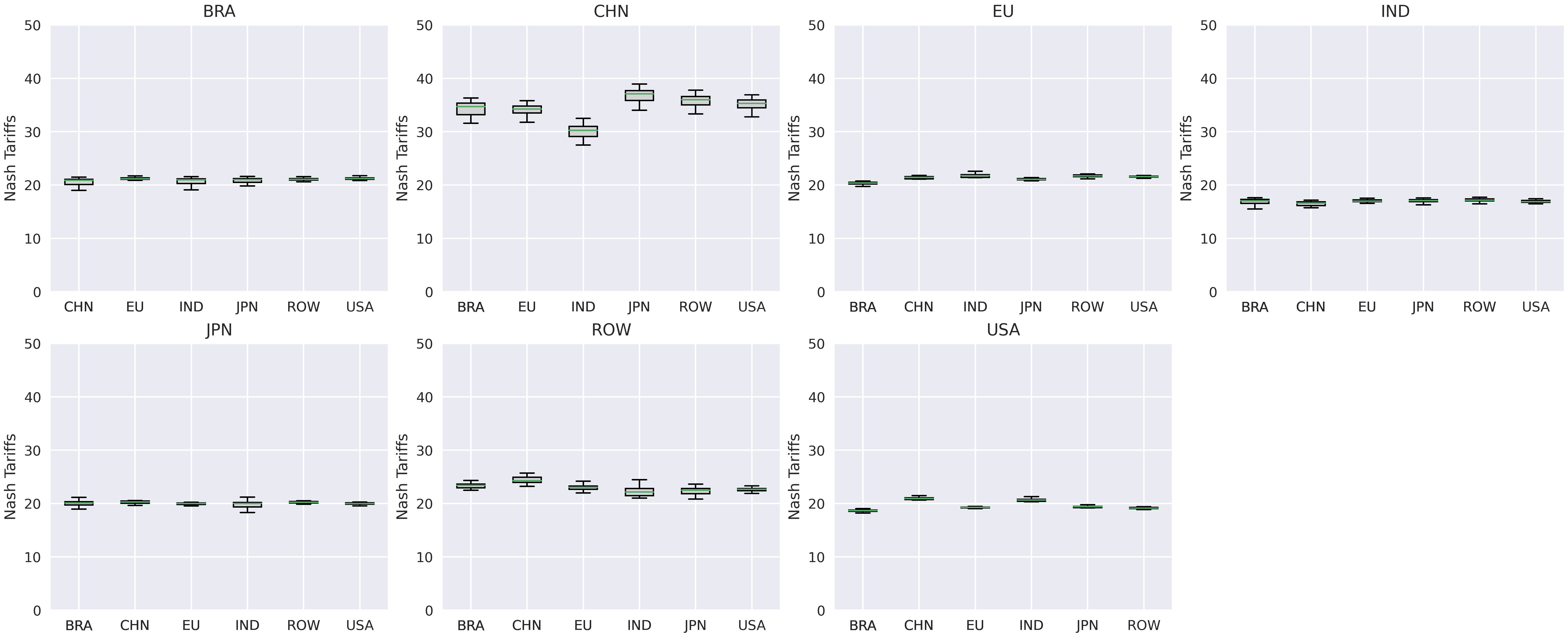}
		
		\caption{Nash Tariffs for Global Dual Policy Competition without Scale Economies}
		\label{fig:nash_tariffs_scale0}
	\end{figure}

	In this subsection, we consider the neoclassical model without scale economies, i.e. $\psi_{j}=0$, as a special case. Figure \ref{fig:nash_tariffs_scale0} illustrates the import tariffs in the Nash equilibrium in which each country decides its tariffs and industrial subsidies to maximize its own real income. It suggests that in a global competition without scale economies, each economy tends to apply almost identical tariffs to all other economies across all industries.
		
\begin{figure}[htbp]
    	\centering
    	\includegraphics[width=0.35\textwidth]{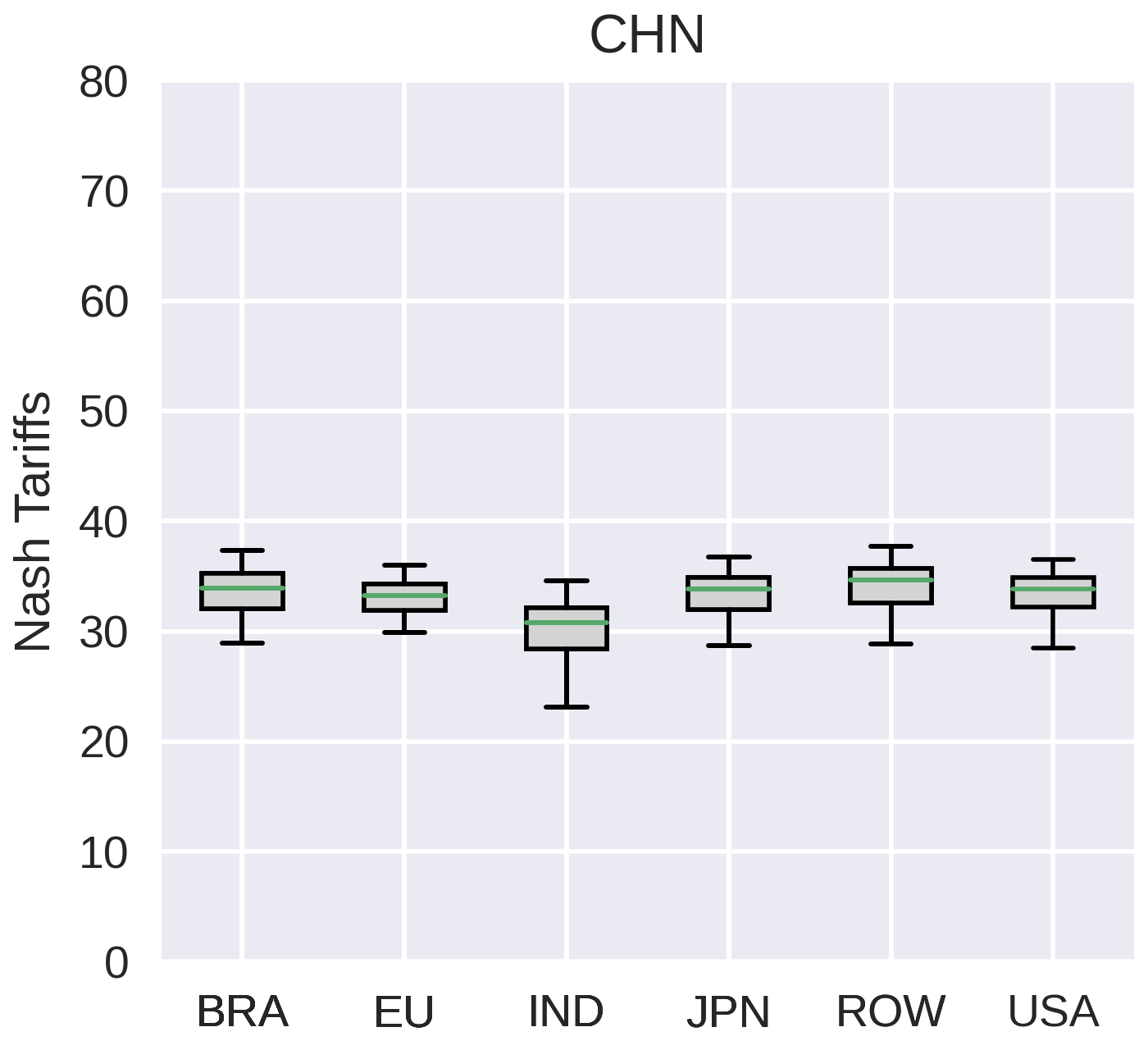}
    	
    	\caption{Import Tariffs in China's Unilaterally Optimal Dual Policies Without Scale Economies}
    	\label{fig:china_uni_dual_scale0_tariff}
    \end{figure}

    Similar to Nash tariffs in Figure \ref{fig:nash_tariffs_scale0}, import tariffs under unilaterally optimal policies, as illustrated by Figure \ref{fig:china_uni_dual_scale0_tariff}, are also almost identical across industries and economies.

	Table \ref{table:nash_wefare_scale0} presents welfare changes under various Nash equilibria. The ``subsidy"	column denotes cases where players can only adjust their industry subsidies. The ``tariff" column signifies
	situations where players can solely modify their import tariffs. Finally, the ``dual" column indicates scenarios
	where players have the flexibility to adjust both their industry subsidies
	and import tariffs.
	
	When considering a competition solely between China and the U.S. without scale economies, the outcomes vary depending on the policy pursued. However, it is important to note that the effects of all three scenarios
	are relatively small.
	
	In the absence of scale economies, the imposition of tariffs and the implementation of dual policies (Columns 5 and 6, respectively) have adverse effects on all seven economies. These competition scenarios give rise to Prisoner's Dilemma situations. When there is only subsidy competition (Column 5), Brazil and China experience slight benefits
	from the competition, while the other economies are negatively impacted.
		
	\begin{table}[htbp]\footnotesize
		\centering
		\caption{Welfare Effects of Nash Policies: $\psi=0$}
		\label{table:nash_wefare_scale0}
		
		\begin{tabular}{lcccccc}
			\toprule 
			 & \multicolumn{3}{c}{\textbf{China and US ($\Delta\%$)}} & \multicolumn{3}{c}{\textbf{World ($\Delta\%$)}}\\\cmidrule(lr){2-4}\cmidrule(lr){5-7}
			& Subsidy & Tariff & Dual & Subsidy & Tariff & Dual\\
			& (1)    &   (2)   & (3)   & (4) & (5) & (6) \\
			Brazil & -0.0264 & 0.0014 & -0.0248 & 0.0102 & -1.1504 & -1.1277\\
			China & 0.1262 & -0.0488 & 0.0713 & 0.0445 & -0.5856 & -0.5247\\
			European Union & -0.0228 & 0.0022 & -0.019 & -0.0492 & -1.0517 & -1.0265\\
			India & -0.0152 & 0.0063 & -0.0057 & -0.0951 & -1.6609 & -1.6261\\
			Japan & -0.0385 & 0.0012 & -0.0368 & -0.0757 & -1.1136 & -1.0731\\
			Rest of the World & -0.1035 & -0.0004 & -0.1039 & -0.0794 & -1.8126 & -1.832\\
			United States & -0.013 & -0.0362 & -0.0467 & -0.0502 & -0.9101 & -0.8904\\
			\bottomrule
		\end{tabular}
	\end{table}

\end{document}